\documentclass[aps,prl,10pt,twocolumn, superscriptaddress, floatsintext, showpacs]{revtex4-2}

\usepackage{graphicx}
\usepackage{lipsum}
\usepackage{adjustbox}
\usepackage{changepage}
\usepackage{float} 
\usepackage{siunitx} 
\usepackage{booktabs} 
\usepackage{amssymb}
\usepackage{amsmath}
\usepackage[colorlinks=true,linkcolor=black,citecolor=blue,filecolor=black,urlcolor=blue]{hyperref}
\usepackage{cleveref}
\usepackage{nth}
\usepackage{gensymb}
\usepackage{xcolor,colortbl}
\usepackage{array}    
\newcolumntype{V}{>{\centering\arraybackslash} m{1.5em}}

\renewcommand{\selectlanguage}[1]{}

\begin{document}

\title{Multi-Orbital Interactions and Spin Polarization in Single Rare-Earth Adatoms}

\author{Massine Kelai}
\affiliation{Center for Quantum Nanoscience (QNS), Institute for Basic Science (IBS), Seoul 03760, Republic of Korea}
\affiliation{Ewha Womans University, Seoul 03760, Republic of Korea}

\author{Stefano Reale}
\affiliation{Center for Quantum Nanoscience (QNS), Institute for Basic Science (IBS), Seoul 03760, Republic of Korea}
\affiliation{Ewha Womans University, Seoul 03760, Republic of Korea}
\affiliation{Department of Energy, Politecnico di Milano, Milano 20133, Italy}

\author{Roberto Robles}
\affiliation{Centro de Física de Materiales CFM/MPC (CSIC-UPV/EHU), Donostia-San Sebastián, Spain}

\author{Jaehyun Lee}
\affiliation{Center for Quantum Nanoscience (QNS), Institute for Basic Science (IBS), Seoul 03760, Republic of Korea}
\affiliation{Department of Physics, Ewha Womans University, Seoul 03760, Republic of Korea}

\author{Divya Jyoti}
\affiliation{Centro de Física de Materiales CFM/MPC (CSIC-UPV/EHU), Donostia-San Sebastián, Spain}
\affiliation{Donostia International Physics Center (DIPC), Donostia-San Sebastián, Spain}

\author{Philippe Ohresser}
\affiliation{Synchrotron SOLEIL, L’Orme des Merisiers, 91190 Saint-Aubin, France}

\author{Edwige Otero}
\affiliation{Synchrotron SOLEIL, L’Orme des Merisiers, 91190 Saint-Aubin, France}

\author{Fadi Choueikani}
\affiliation{Synchrotron SOLEIL, L’Orme des Merisiers, 91190 Saint-Aubin, France}

\author{Fabrice Scheurer}
\affiliation{Institut de Physique et Chimie des Matériaux de Strasbourg, Strasbourg, France}

\author{Nicol\'as Lorente}
\affiliation{Centro de Física de Materiales CFM/MPC (CSIC-UPV/EHU), Donostia-San Sebastián, Spain}
\affiliation{Donostia International Physics Center (DIPC), Donostia-San Sebastián, Spain}

\author{Deung-Jang Choi}
\affiliation{Centro de Física de Materiales CFM/MPC (CSIC-UPV/EHU), Donostia-San Sebastián, Spain}
\affiliation{Donostia International Physics Center (DIPC), Donostia-San Sebastián, Spain}

\author{Aparajita Singha}
\affiliation{Max Planck Institute for Solid State Research, Stuttgart, Germany}

\author{Fabio Donati}
\email{Corresponding author. Email: donati.fabio@qns.science}
\affiliation{Center for Quantum Nanoscience (QNS), Institute for Basic Science (IBS), Seoul 03760, Republic of Korea}
\affiliation{Department of Physics, Ewha Womans University, Seoul 03760, Republic of Korea}

\begin{abstract}
Surface-adsorbed rare-earth nanostructures are ideal platforms to investigate the interplay between intra-atomic interactions and multi-orbital spin configurations. However, addressing these properties has posed severe experimental and theoretical challenges. Here, we use the orbital selectivity offered by X-ray absorption spectroscopy to quantify the Coulomb integrals of Nd atoms on conductive surfaces, as well as the variation of individual orbital occupation upon cluster nucleation. Using X-ray magnetic circular dichroism we identify magnetic moments of the order of few tens of~$\mu_{\rm{B}}$ at the $5d$ orbitals and their magnetic coupling with the $4f$ spins. Our results validate orbital-resolved X-ray spectroscopy as a reliable method for quantifying complex multi-orbital interactions in surface-adsorbed lanthanides. 
\end{abstract}

\maketitle
Rare-earth (RE) elements represent exquisite building blocks to realize materials with various and intriguing properties such as frustrated magnetism~\cite{greedan_frustrated_2006, lago_cder_2010}, superconductivity~\cite{sun_pressure-induced_2010, yamaoka_role_2014}, heavy fermion~\cite{rueff_pressure-induced_2011}, and mixed-valence behavior~\cite{batlogg_valence_1981, matsumura_pressure_1997, grazioli_characteristic_2001, butch_pressure-resistant_2016, joseph_experimental_2017}. The intricate interplay between strongly localized $\textrm{4}f$ orbitals and more delocalized $\textrm{5}d$ and $\textrm{6}s$ valence orbitals plays a major role in influencing ultrafast spin dynamics~\cite{wietstruk_hot-electron-driven_2011, radu_transient_2011, khorsand_element-specific_2013, bergeard_ultrafast_2014, frietsch_disparate_2015}, complex magnetic interactions~\cite{radu_transient_2011, bergeard_ultrafast_2014, peters_magnetism_2016}, and it is pivotal in both shaping the magnetic anisotropy and enhancing the magnetic relaxation of atomic and molecular magnets~\cite{luis_molecular_2011, aguila_heterodimetallic_2014, meihaus_record_2015, liu_single_2017, gould_synthesis_2019, dubrovin_magnetism_2019, dubrovin_valence_2021}.

In the last decade, individual RE atoms on surfaces have been explored as a model system for ultra-high-density magnetic data storage~\cite{baltic_superlattice_2016, donati_magnetic_2016, singha_magnetic_2016, natterer_reading_2017, donati_correlation_2021}, spin qubit candidates~\cite{reale_erbium_2023}, as well as platforms for the emergence of exotic many-body states~\cite{yazdani_probing_1997, ding_tuning_2021}. For these systems, the presence of $5d6s$ valence electrons can significantly enhance the spin contrast in transport measurement~\cite{pivetta_measuring_2020, curcella_valence_2023} and the coupling with the substrate electrons~\cite{coffey_antiferromagnetic_2015}. Although theoretical approaches have been widely used to shed light on the multi-orbital configuration of RE adatoms~\cite{coffey_antiferromagnetic_2015, pivetta_measuring_2020, donati_correlation_2021}, it is challenging to experimentally identify the multi-orbital configuration for RE adsorbed on surfaces, and it has been demonstrated only for Gd and Ho atoms on MgO/Ag(100)~\cite{singha_mapping_2021}.

Previous studies found two distinct $4f$ occupations for RE adatoms, namely, atomic-like $4f^{n}$ and bulk-like $4f^{n-1}$, established through a balance between intra-atomic interactions, coordination, and RE-substrate hybridization~\cite{nistor_structure_2014, singha_4_2017, baltic_magnetic_2018, donati_correlation_2021}. The determining factors for $4f$ occupation and spin configuration of the ground state are the intra- and inter-orbital Coulomb integrals, as well as the occupation and polarization of the outer $5d6s$ shells~\cite{kotani_many-body_1988, bianconi_localization_1982, baudelet_study_1993}. However, accurately computing these quantities has been a long-standing problem for theoretical methods attempting to predict the many-body ground state of strongly correlated systems containing REs~\cite{gunnarsson_electron_1983, kotani_many-body_1988, van_der_marel_electron-electron_1988, malterre_l_1991, vaugier_2011, jiang_electronic_2012, amadon_screened_2014, lanata_phase_2015, liu_comparative_2023}, calling for experimental approaches to advance the understanding of these materials.

In this study, we employ X-ray absorption spectroscopy (XAS) to characterize the multi-orbital configuration of Nd adatoms. On weakly interacting substrates such as HOPG and Ag(100), Nd adatoms exhibit an atomic-like $4f^{4}$ occupation, transitioning to bulk-like $4f^{3}$ occupations upon adsorption on strongly interacting substrates [Pb(111)] or with increased coordination through cluster nucleation. Using orbitally-resolved X-ray transitions along the $M_{2,3,4,5}$ series~\cite{singha_mapping_2021} we find substrate-dependent variations of the absorption spectra, which allow us to quantify both the Coulomb integrals and variations in individual shell occupation upon cluster nucleation. X-ray magnetic circular dichroism (XMCD) unravels the spin polarization of Nd atoms and clusters with orbital sensitivity, together with their characteristic inter-orbital magnetic coupling. For the $5d$ orbitals, we find a magnetic moment of few tens of~$\mu_{\rm{B}}$, in agreement with density functional theory (DFT). Our results confirm the capability of the proposed method in addressing the complexity of RE surface-supported nanostructures.

\begin{figure}[t]
        \centering \includegraphics[width=0.47\textwidth, trim=0.25 0 0 0.5 cm, clip]{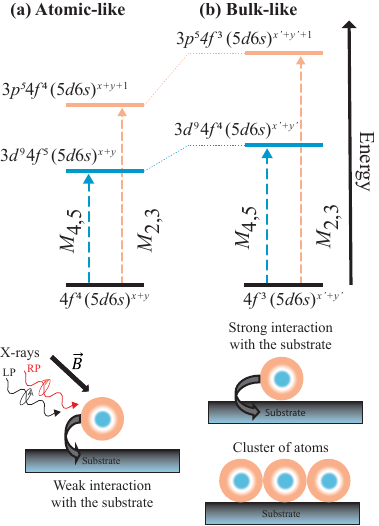}
        \caption{XAS transitions diagram for Nd on: (a) a weakly interacting substrate (atomic-like: $\textrm{4}f^{4}$); (b) a strongly interacting substrate or in cluster form (bulk-like: $\textrm{4}f^{3}$). $M_{5}$ (blue) and $M_{3}$ (salmon) probe the $\textrm{4}f$ and $\textrm{5}d\textrm{6}s$ occupancies, respectively. Energy differences denoted highlight transitions involving different electron arrangements in the $\textrm{4}f$ and $\textrm{5}d\textrm{6}s$ orbitals. $x$ (resp. $x'$) and $y$ (resp. $y'$) represent the number of electrons in the $\textrm{5}d$ and \textrm{6}s orbitals in the atomic-like (resp. bulk-like) ground state configuration. The magnetic field vector ($\vec{B}$) and the wave vector of circularly polarized photons are parallel.}
        \label{fig:1}
        \vspace{-0.5cm}
\end{figure}

XAS and XMCD measurements were conducted at the DEIMOS beamline~\cite{ohresser_deimos_2014, Joly_2014, Joly_2016} of the SOLEIL synchrotron. These measurements involved taking the sum and difference of right and left circular polarization absorptions under a magnetic field  $B=6$\,T and a temperature $T=2$\,K. To ensure the stabilization of individual adatoms and prevent surface diffusion, we sublimated minute amounts of Nd (less than 1\% of a monolayer, ML) onto the sample kept below 10~K, while larger Nd coverages were employed to promote the statistical growth of clusters~\cite{brune_microscopic_1998}. Surface and orbital sensitivity were achieved through soft X-ray absorption at specific $M_{\alpha}$ edges. The electric dipole selection rules impose a unitary change in the total orbital quantum number. Consequently, the $M_{4,5}$ transitions create a hole in the $\textrm{3}d$ core levels by transferring one electron to $\textrm{4}f$, while the $M_{2,3}$ transitions promote one electron from the $\textrm{3}p$ to the $\textrm{5}d\textrm{6}s$ states, as illustrated in Fig.~\ref{fig:1}. As demonstrated further in this paper, each transition provides complementary information to map the multi-orbital configuration of RE atoms.

\begin{figure}[t]
\centering
\includegraphics[width=0.49\textwidth, trim=0.5 0.5 0 0cm, clip]{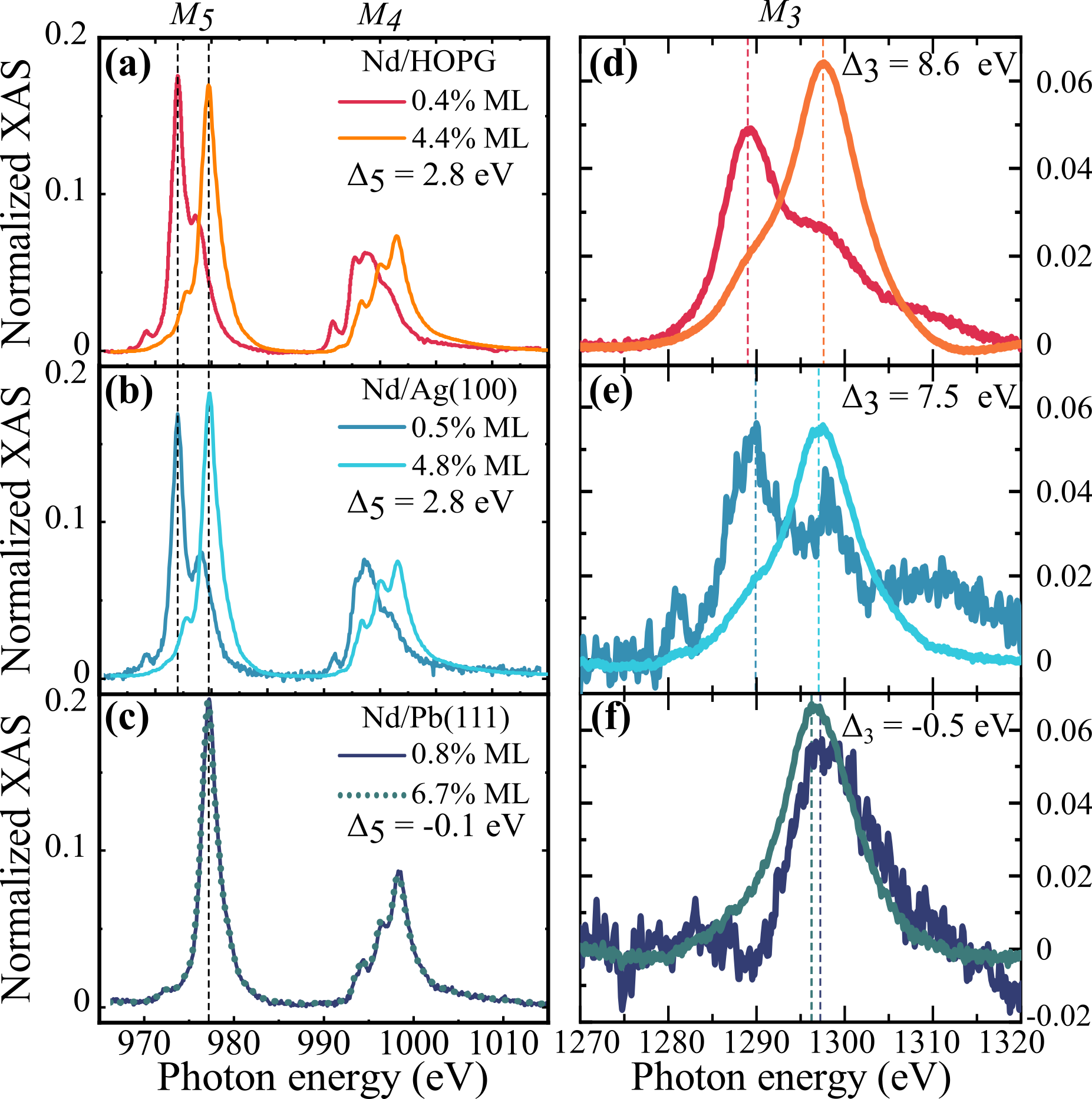}
\caption{X-ray absorption of Nd on HOPG, Ag(100), and Pb(111) at the $M_{4,5}$ edge (a, b, c) and $M_{3}$ edge (d, e, f). Dashed lines mark the XAS maxima for each sample. The energy differences between the atoms' and clusters' XAS maxima at the $M_{5}$ and $M_{3}$ edges are marked as $\Delta_{5}$ and $\Delta_{3}$, respectively. The Nd coverage in \%~ML is indicated in the legend. The measurements have been acquired at 6~T and 2~K along the easy magnetization axis for isolated Nd atoms on each substrate: normal incidence on HOPG and Ag(100), and grazing incidence on Pb(111)~\cite{SI}.}
\label{fig:2}
\vspace{-0.35cm}
\end{figure}

Similar to other REs~\cite{singha_4_2017}, the choice between atomic-like $4f^{4}$ and bulk-like $4f^{3}$ occupations can be influenced by the atom-substrate interaction and coordination. The multi-orbital configuration of RE atoms leads to specific features in the XAS, whose energy positions are different at the $M_{4,5}$ and $M_{2,3}$ edges because of different core hole-valence interactions. As illustrated in Figs.~\ref{fig:2}a-c, on HOPG and on Ag(100), the $M_{5}$ XAS of isolated Nd atoms distinctly reflects a dominant $4f^{4}$ atomic-like configuration~\cite{baltic_magnetic_2018}, while on Pb(111), the XAS maximum is observed at a higher energy, with an absorption line indicating a $4f^{3}$ bulk-like configuration~\cite{thole_3d_1985}. This difference in $4f$ occupation suggests a weaker atom-substrate interaction on HOPG and Ag(100) compared to Pb(111)~\cite{singha_4_2017}. To understand the substrate-dependence of $4f$ occupation, we compare our findings with DFT calculations performed using the generalized gradient approximation for the exchange-correlation functional (GGA+U) and employing a plane-wave basis and PAW potentials~\cite{SI}. For HOPG and Ag(100), we find Nd atoms to exhibit preferential adsorption at the hollow site and $4f^{4}$ occupation, while on the Pb(111) surface embedding is favored~\cite{SI}. The increased coordination at the equatorial atomic plane promotes a $4f^{3}$ configuration and an opposite sign of the magnetic anisotropy on this substrate~\cite{singha_magnetic_2016}. 

Upon cluster nucleation, realized by increasing the Nd coverage, the maxima of the $M_{5}$ XAS of both HOPG and Ag(100) move up in energy by $\Delta_{5} =~2.8$~eV, and align with the maximum observed on Pb(111), indicating a transition from $4f^{4}$ to $4f^{3}$ occupation. We also find similar behavior for Nd on graphene(Gr) on Cu(111)~\cite{SI}. This energy difference is also evident at the $M_{4}$ edge~\cite{SI} and is comparable to values reported in the literature for various RE atoms~\cite{singha_4_2017, donati_correlation_2021}. In contrast, on Pb(111), the spectrum shows a negligible redshift upon cluster nucleation ($\Delta_{5}=-0.1$~eV), suggesting no significant changes in the $4f$ occupation. The scenario becomes more diversified for the spectra acquired at the $M_{3}$ edge, as presented in Figs.~\ref{fig:2}d-f. Again on HOPG and Ag(100), the XAS maximum of the ensemble of isolated atoms moves to higher energy upon cluster nuclation. In this case, however, the energy difference is notably larger than that observed for the $M_{5}$ edge and displays variations between the two substrates, i.e. $\Delta_{3} = 8.6$~eV and $7.5$~eV for HOPG and Ag(100), respectively. A similar trend is also observed at the $M_{2}$ edge~\cite{SI}. In contrast, Nd on Pb(111) exhibits a much smaller and opposite shift of the XAS maximum ($\Delta_{3}=-0.5$~eV). These findings lead us to conclude that on both $M_{5}$ and $M_{3}$ edges the position of the XAS maxima is mostly determined by the $\textrm{4}f$ occupation. However, the larger energy differences observed at the $M_{3}$ edge allow detecting sizable substrate-dependent shifts, which offers the possibility to quantify changes in the electronic configuration of the outer shells. 

To this extent, we developed a multi-orbital analytical model to compute the quantities $\Delta_{5}$ and $\Delta_{3}$. In its simplest formulation, the model includes only $4f$ and $5d$ orbitals, although it can be generalized to include an arbitrary set of orbitals for any XAS transition~\cite{SI}. In this two-orbital model, the energy difference of each transition $\Delta_{5}$ and $\Delta_{3}$ can be expressed as:
\vspace*{-1mm}
\begin{subequations}
    \begin{align}
    \label{eq:M45}
    \Delta_{5} &= \delta \textrm{q}_{4f}(U_{4f4f} - U_{3d4f}) 
    + \delta \textrm{q}_{5d}(U_{4f5d} - U_{3d5d}) \\
    \label{eq:M23}    
    \Delta_{3} &= \delta \textrm{q}_{4f}(U_{4f5d} - U_{3p4f}) 
    + \delta \textrm{q}_{5d}(U_{5d5d} - U_{3p5d})
    \end{align}
\end{subequations}

\noindent where $\delta \textrm{q}_{4f,5d}$ refers to the difference between the cluster and atom occupations at the $4f$ and $5d$ orbitals, respectively, and $U_{\alpha,\beta}$ are the intra- and inter-shell Coulomb integrals. Although Eqs.~\eqref{eq:M45} and \eqref{eq:M23} are exact for integer charges variations, the linear response of the interaction energy with the orbital occupation ensures their validity also with fractional charges~\cite{cococioni_linear_2005}.  In both equations, the terms are arranged in increasing values of the Coulomb integrals. 

All terms contain a difference between the multi-orbital Coulomb integrals involving core states and another one involving only $4f$ or $5d$ states. As the latter always overcomes the former~\cite{SI}, these differences are always negative, hence the sign of the individual terms depends on $\delta \textrm{q}_{4f,5d}$. Thus, when a change in $4f$ occupation occurs between atoms and clusters ($\delta \textrm{q}_{4f} = -1$), as observed for Nd on HOPG and Ag(100), the first terms provide a large positive contribution. For these systems, the change of $4f$ occupation upon cluster formation needs compensation through an opposite change in $5d$ occupation ($\delta \textrm{q}_{5d} > 0$), hence the second terms provide a negative contribution, although smaller in magnitude than the first terms. For Nd on Pb(111), the absence of changes in $4f$ occupation makes the first term of both equations null, and the related negative energy shifts $\Delta_{5}$ and $\Delta_{3}$ are only due to negative contributions of the second term, in line with the experiment. This observation also indicates a positive change of $\delta \textrm{q}_{5d}$ upon cluster formation even without changes in the $4f$ occupation, possibly due to differences in charge transfer to the substrate between atoms and clusters.
\begin{table}[b]
    \vspace{-0.8 cm}
    \centering
    \caption{Charge variation for Nd on different substrates obtained from the fit to the data using Eqs.~\eqref{eq:M45} and \eqref{eq:M23}. The orbital-dependant of screening coefficients are shown in the two last columns.}
        \begin{tabular}{lcccc || cc}
        \toprule
        Nd on & {HOPG} & {Gr/Cu(111)} & {Ag(100)} & {Pb(111)} & $\alpha$ & $\kappa_{\alpha}$ \\
        \midrule
        $\delta \textrm{q}_{4f}$ & -1 & -1 & -1 & 0 &  $3p,3d,4f$ & 0.29 \\
        $\delta \textrm{q}_{5d}$ & 0.50 & 0.23 & 1.22 & 0.45 & $5d$ & 0.06\\
        \bottomrule
    \end{tabular}
    \label{tab:q&k}
\end{table}

To quantify the changes in orbital occupation, we fit the values of $\Delta_{5}$ and $\Delta_{3}$ using Eqs.~\eqref{eq:M45}~and~\eqref{eq:M23} through a Bayesian optimization procedure~\cite{frazier_tutorial_2018, SI}. To account for the electron-electron screening from the substrate, the Coulomb integrals are computed as $U_{\alpha,\beta}~=~\sqrt{\kappa_{\alpha}\kappa_{\beta}} F^0_{\alpha,\beta}$, where $\kappa_{\alpha,\beta}$ are effective screening coefficients and $F^0_{\alpha,\beta}$ are the first-order Slater integrals obtained from a Hartree-Fock atomic calculation~\cite{cowan_theory_nodate}. To avoid overparametrization, we assume $\kappa_{3p}=\kappa_{3d}=\kappa_{4f}$, as these orbitals are not significantly  hybridized with the substrate, while we take $\kappa_{5d}$ as an independent value. This reduces the number of free-parameters related to the $U_{\alpha,\beta}$ to two values. In addition, we fix the value of $\delta \textrm{q}_{4f}$ from the energy shift of the $M_{5}$ edge and we obtain $\delta \textrm{q}_{5d}$ from the fit (one for each substrate), under the assumption of a cluster-size independent charge transfer. We constraint the fit by imposing charge conservation upon cluster nucleation,  
i.e., $\delta \textrm{q}_{4f} + \delta \textrm{q}_{5d} + \delta \textrm{q}_{\textrm{env}} = 0$, where 
the environment charge variation ($\delta \textrm{q}_{\textrm{env}}$) describes the charge transfer towards the $6s$ orbitals, as well as to the substrate. We constrain the latter term between two extreme cases, namely, $-0.5 \leq \delta \textrm{q}_{\textrm{env}} \leq - \delta \textrm{q}_{\textrm{4f}}$. The lower bound represents the maximum back donation from the environment, quantified from typically observed charge variation in RE~\cite{singha_mapping_2021, nistor_structure_2014} and alkaline adatoms~\cite{kovaric_alkaline_2022} upon cluster nucleation. Instead, the upper bound corresponds to the case where all the $4f$ charge is transferred towards the environment, without altering the $5d$ occupation.

\begin{figure*}[t]
\centering
\includegraphics[width=0.99\textwidth, keepaspectratio]{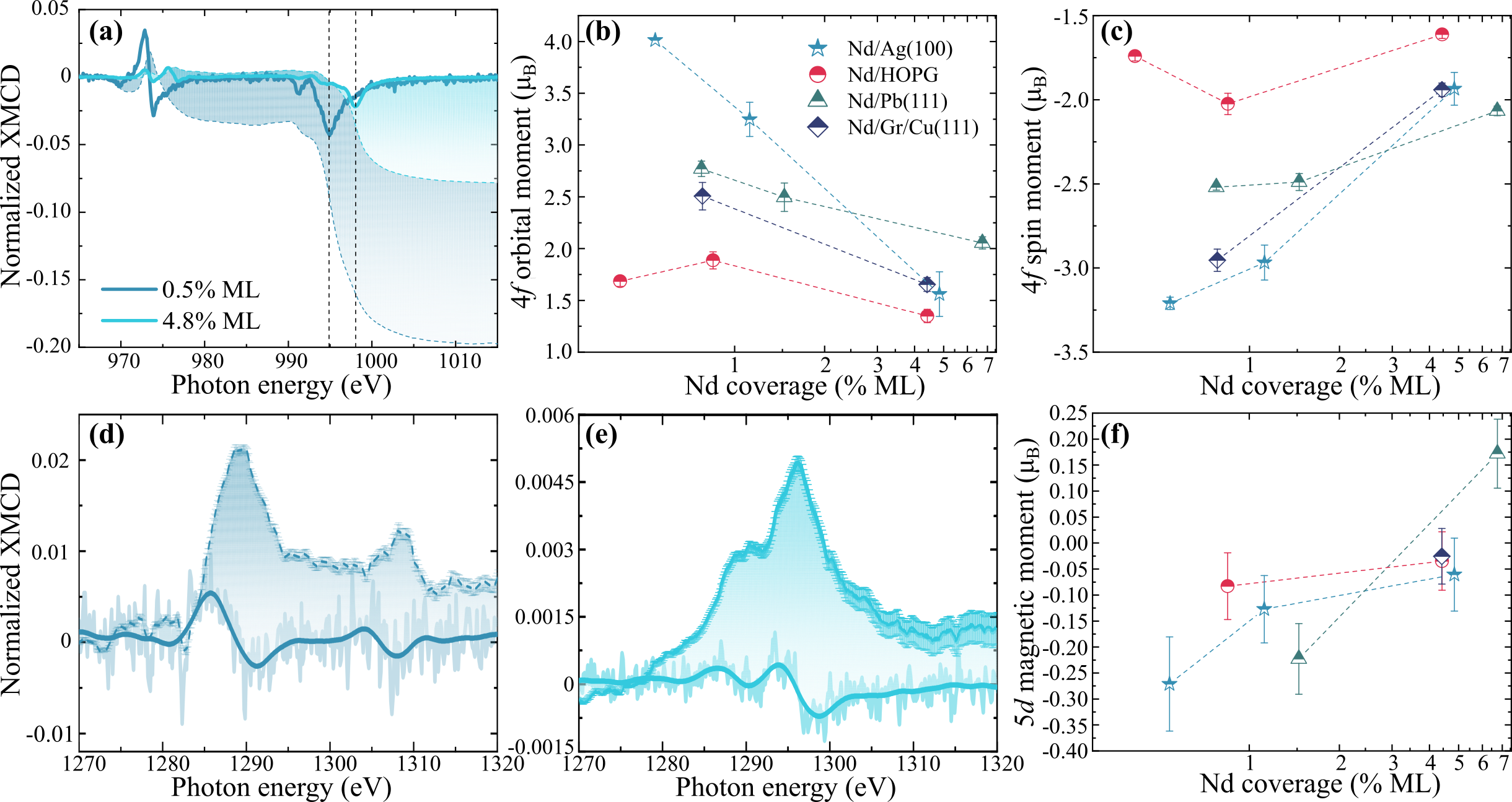}
\caption{Normalized and integrated XMCD spectra at the $M_{4,5}$ edges for Nd on Ag(100) are presented for the lowest (in blue) and highest (in cyan) coverages (a). Dashed vertical lines denote the energy corresponding to the XMCD maximum for both configurations. $4f$ orbital and spin moments, obtained through sum rules, are shown in (b) and (c). Normalized and integrated XMCD spectra at the $M_{3}$ edge for individual atoms and clusters are depicted in (d) and (e), respectively. Extracted magnetic moments are highlighted in panel (f). All $M_{3}$ XMCD spectra are normalized to the integral of the $M_{2,3}$ edge ($t_{M_{2,3}}$) shown in Tab.~S\textrm{III}~\cite{SI}, and the $p_{M_{3}}$ integral value is measured at 1320~eV. Raw data, displayed in transparent lines, are smoothed using a Sawatzky-Golay method (full lines) for enhanced clarity. The error bar is represented by the deviation around the central line of the XMCD integral (dashed lines). All the data were acquired at 6~T   and 2~K along the easy magnetization axis for isolated Nd atoms on each substrate~\cite{SI}.}
\label{fig:3}
\vspace{-0.5 cm}
\end{figure*}

The charge variations obtained from the fit exhibit clear substrate dependence. On HOPG and Gr/Cu(111), the charge coming from the $4f$ is partially transferred to the $5d$. This implies that a significant fraction of the atomic $4f$ charge is transferred to the environment, namely, the $6s$ orbitals and the substrate. This tendency can be explained by considering the electronic structure of RE atoms on graphene with open $6s$ shells \cite{pivetta_measuring_2020, curcella_valence_2023}, which may preferentially get filled upon bond formation. Conversely, on Ag(100) and Pb(111), the $5d$ orbitals gather more charge than what is transferred from the $4f$, indicating back donation from the environment. Please note that similar charge transfers are obtained in a three-orbital model including the $6s$ orbitals~\cite{SI}. Variations in $\Delta_{3}$ across different samples are hallmarks of substrate-dependent charge transfer. Orbital-resolved quantification of the charge variations occurring upon nucleation of surface clusters is a unique capability of the proposed approach, which relies solely on XAS acquisition at the $M_{\alpha}$ edges.

The proposed model finds additional validation in the comparison between the obtained values of the Coulomb integrals and those reported in the literature for bulk samples. We find $U_{4f4f}=7.8$~eV, which closely resembles the value reported for metallic Nd from $3d$ X-ray photoemission spectroscopy~\cite{lang_study_1981, van_der_marel_electron-electron_1988}. Similarly, $U_{4f5d} = 1.6$~eV falls within the range reported for various lanthanum compounds \cite{kotani_many_1987}. We notice that the screened Coulomb integrals involving the $6s$ orbitals from a three-orbital model are negligible compared to $U_{\alpha,4f}$ and $U_{\alpha,5d}$, thus supporting the use of a two-orbital model~\cite{SI}.

The use of XMCD further allows us to address the individual orbital spin polarization of RE atoms and their variation occurring upon nucleation. Fig.~\ref{fig:3}a shows the XMCD of Nd atoms and clusters at the $M_{4,5}$ edges on Ag(100), which probes the magnetism of the $4f$ electrons. Among the investigated samples, Nd atoms on Ag(100) show the largest XMCD, which is maximal at normal incidence~\cite{SI}, i.e., out-of-plane magnetic easy axis. As typically observed for an early lanthanide, the largest dichroism is found at the $M_{4}$ edge~\cite{schille_4_1994, baltic_magnetic_2018}. Upon cluster nucleation, the XMCD peak shifts at higher energy by the same amount found for the XAS in Fig.~\ref{fig:2}b owing to the change in $4f$ occupation. At the same time, its intensity decreases due to either a change of the magnetic easy axis~\cite{singha_magnetic_2016, SI} or to the formation of non-collinear magnetic states~\cite{van_dijk_unusual_2015}, both preventing reaching saturation at the maximum available field (6~T). Note that in the case of the lowest coverage we can assume that the XMCD spectrum is dominated by the magnetic contribution of the isolated atoms. The same can be said for the highest coverage, as we only have a collection of large clusters. However, in the case of intermediate coverage, the system consists of a mixture of adatoms and clusters of different sizes~\cite{brune_microscopic_1998}, so that the XMCD signal is actually a mixture of signals emanating from different species~\cite{mocuta_artificial_2021}. Within these limitations, we can still follow the trend of the estimated magnetic moments over the ensemble. As shown in Fig.~\ref{fig:3}b/c, for all substrates except HOPG, sum rules~\cite{thole_x-ray_1992, carra_x-ray_1993} reveal the large magnetic moments for the lowest coverage of Nd, with an opposite direction between the spin ($\mu_{S}^{4f}$) and the orbital ($\mu_L^{4f}$) magnetic moments, as expected for an early lanthanide with less than half-filled $4f$ shells. Similarly to Nd on Ag(100), a decreasing trend of magnetic moments with increasing Nd coverage is evident for all samples except HOPG regardless of their anisotropy, hence suggesting the formation of a canted magnetic ordering within clusters~\cite{gerion_high_1999, van_dijk_unusual_2015, SI}. The behavior found on HOPG can be ascribed to a smaller magnetic moment of the $4f^{4}$ species on this substrate compared to that of the $4f^{3}$ ones whose abundance increases with the Nd coverage~\cite{SI}.

\begin{table}[t]
\vspace{-0.2 cm}
\centering
\caption{Magnetic moments ($\mu_{5d}$) and occupation ($\textrm{q}_{5d}$) of the $5d$ orbitals of Nd. Magnetic moments are given in Bohr magnetons $\mu_B$. Uncertainty on the last digit is indicated in bracket.}
\begin{tabular}{l S[table-format=1.5] S[table-format=1.5]  || S[table-format=1.2] S[table-format=1.2]}
\toprule
 & {XMCD} & & {DFT} \\
\midrule
 Coverage (ML) & {$0.4\%$} & {$0.8-1.6\%$} & {\hspace{0 cm} Single atom}  \\
\midrule
 Nd on & {$\lvert \mu_{5d} \rvert$} & {$ \lvert \mu_{5d} \rvert$} & {$\mu_{5d}$} & {$\textrm{q}_{5d}$} \\
\midrule
HOPG & {---} & 0.08\pm0.06    & 0.09 & 0.24 \\
Ag(100) & 0.27\pm0.09 & 0.13\pm0.07 & 0.06 & 0.27 \\
Pb(111) & {---} & 0.22\pm0.07 & 0.17 & 0.83 \\
\bottomrule
\end{tabular}
\label{tab_m}
\vspace{-0.5 cm}
\end{table}

The XMCD acquired at the $M_{3}$ edge (Fig.~\ref{fig:3}d/e) exhibits a characteristic peak-derivative shape for Nd atoms and clusters on Ag(100), as well as on all other substrates~\cite{SI}. The peak-derivative reverses its sign at the $M_{2}$ edge~\cite{SI}, similar to the behavior observed for Gd on MgO/Ag(100)~\cite{singha_mapping_2021}. A comparison with multiplet calculations allows us to conclude that the XMCD signal is largely dominated by transitions to the $5d$ states even for systems with an open $6s$ shell, possibly due to stronger renormalization of the $6s$ Slater integrals~\cite{SI}. The peak-derivative shape of Nd XMCD shows an opposite sign compared to Gd~\cite{singha_mapping_2021}. This reversed XMCD shape arises from the opposite alignment between the $5d$ spin of Gd and Nd with respect to the magnetic field. While in Gd both $4f$ and $5d$ spins align along the external field due to the vanishing orbital moment, in Nd the alignment along the magnetic field is determined by the $4f$ orbital moment due to its largest magnitude. In turn, the spin-orbit coupling induces the $4f$ spins and the exchange-coupled $5d$ spins to align opposite to the field, leading to reversed $M_{2,3}$ XMCD.

The relative small contribution of the $4f-5d$ exchange-induced \textit{breathing}~\cite{van_veenendaal_branching_1997, parlebas_x-ray_2006} to the $M_{2,3}$ XMCD~\cite{singha_mapping_2021, SI} allows us to apply sum rules to the spectra~\cite{thole_x-ray_1992, carra_x-ray_1993} and extract the values of the $5d$ magnetic moments. As shown in Fig.~\ref{fig:3}d/e, a positive value of the $M_{3}$ XMCD integral is observed for atoms and clusters on Ag(100), as well as for all samples, while the lowest intensity of the $M_{2}$ XMCD often leads to large uncertainty in the value of this integral~\cite{SI}. The impact of this uncertainty on the sum rules can be reduced by computing the total $5d$ magnetic moment $\mu_{5d}$~\cite{SI}, whose values are summarized in Fig.~\ref{fig:3}f. This magnetic moment has the same sign of $\mu_{\textrm{S}}^{4f}$ and follows the same trend with the coverage (Fig.~\ref{fig:3}c), confirming the parallel $4f$-$5d$ spin coupling. The magnitude of $\mu_{5d}$ is much smaller than $\mu_{\textrm{S,L}}^{4f}$ but in line with values from ferromagnetic RE bulk compounds~\cite{jo_4_1993}. 

As shown in Table~\ref{tab_m}, the results from sum rules are slightly larger but close in magnitude to the values from DFT. This discrepancy may arise due to the well-known shortcoming of DFT in capturing exchange and correlation effects in highly-correlated materials such as RE. For Nd on Ag(100) and HOPG the low polarization originates from the small $5d$ occupation, while for Pb(111) it stems from the large overlap between the spin-split $5d$ states~\cite{SI}.

Our approach based on orbital-resolved absorption spectroscopy provides crucial insights into multi-orbital Coulomb interactions among isolated atoms on surfaces, allowing quantification of changes in the charge state during cluster nucleation. These findings offer an experimental foundation for theoretical models in simulating strongly correlated atoms, addressing the issue of the computational complexities and costs. Moreover, by probing the spin polarization in $4f$ and $5d$ orbitals, our method allows complementing scanning probe techniques that are more sensitive to the spatially-extended $6s$ electronic states~\cite{pivetta_measuring_2020, curcella_valence_2023}, opening the way to a full experimental characterization of surface RE magnetism.

\section{Acknowledgement}
Experiments were performed on the DEIMOS beamline at SOLEIL Synchrotron, France (proposal numbers  20191894 and 20220066). We are grateful to the SOLEIL staff for smoothly running the facility. The authors would like to thank Bernard Muller for his significant contribution to the construction of DEIMOS beamline (design and engineering) and Florian Leduc for his support. The MBE chamber used during the XAS/XCMD experiment on DEIMOS has been funded by the Agence National de la Recherche; grant ANR-05-NANO-073.



\clearpage

%

\end{document}


\title{Supplemtary Information: Multi-Orbital Interactions and Spin Polarization in Single Rare-Earth Adatoms}

\author{Massine Kelai}
\affiliation{Center for Quantum Nanoscience (QNS), Institute for Basic Science (IBS), Seoul 03760, Republic of Korea}
\affiliation{Ewha Womans University, Seoul 03760, Republic of Korea}

\author{Stefano Reale}
\affiliation{Center for Quantum Nanoscience (QNS), Institute for Basic Science (IBS), Seoul 03760, Republic of Korea}
\affiliation{Ewha Womans University, Seoul 03760, Republic of Korea}
\affiliation{Department of Energy, Politecnico di Milano, Milano 20133, Italy}

\author{Roberto Robles}
\affiliation{Centro de Física de Materiales CFM/MPC (CSIC-UPV/EHU), Donostia-San Sebastián, Spain}

\author{Jaehyun Lee}
\affiliation{Center for Quantum Nanoscience (QNS), Institute for Basic Science (IBS), Seoul 03760, Republic of Korea}
\affiliation{Department of Physics, Ewha Womans University, Seoul 03760, Republic of Korea}

\author{Divya Jyoti}
\affiliation{Centro de Física de Materiales CFM/MPC (CSIC-UPV/EHU), Donostia-San Sebastián, Spain}
\affiliation{Donostia International Physics Center (DIPC), Donostia-San Sebastián, Spain}

\author{Fadi Choueikani}
\affiliation{Synchrotron SOLEIL, Gif sur Yvette, France}

\author{Edwige Otero}
\affiliation{Synchrotron SOLEIL, Gif sur Yvette, France}

\author{Philippe Ohresser}
\affiliation{Synchrotron SOLEIL, Gif sur Yvette, France}

\author{Fabrice Scheurer}
\affiliation{Institut de Physique et Chimie des Matériaux de Strasbourg, Strasbourg, France}

\author{Nicol\'as Lorente}
\affiliation{Centro de Física de Materiales CFM/MPC (CSIC-UPV/EHU), Donostia-San Sebastián, Spain}
\affiliation{Donostia International Physics Center (DIPC), Donostia-San Sebastián, Spain}

\author{Deung-Jang Choi}
\affiliation{Centro de Física de Materiales CFM/MPC (CSIC-UPV/EHU), Donostia-San Sebastián, Spain}
\affiliation{Donostia International Physics Center (DIPC), Donostia-San Sebastián, Spain}

\author{Aparajita Singha}
\affiliation{Max Planck Institute for Solid State Research, Stuttgart, Germany}

\author{Fabio Donati}
\email{Corresponding author. Email: donati.fabio@qns.science}
\affiliation{Center for Quantum Nanoscience (QNS), Institute for Basic Science (IBS), Seoul 03760, Republic of Korea}
\affiliation{Department of Physics, Ewha Womans University, Seoul 03760, Republic of Korea}

\maketitle 

\section{1. Model for the X-ray absorption energy shift induced by configurational changes from atoms to clusters}

\subsection{1.1 $M_{4,5}$ XAS transition: $3d^{10}4f^{n}$ $\rightarrow$ $3d^{9}4f^{n+1}$}
The ground state configuration is represented as $3d^{10}4f^{n}5d^{m}6s^{k}$. The excited state is reached by following the dipole selection rules. In the soft X-ray regime, the absorption of a single photon occurs, promoting an electron from the $3d$ core levels to the $4f$ shell, resulting in the excited state configuration $3d^{9}4f^{n+1}5d^{m}6s^{k}$. Thus, in the following, we will calculate the ground state ($E_{GS}$) and excited state ($E_{ES}$) energies:
\begin{subequations}
\begin{equation}\label{M45nGS} 
\begin{split}
 E_{GS}(n,m,k) &= 10\varepsilon_{3d} + n \varepsilon_{4f} + \frac{n(n-1)}{2} U^{GS}_{4f4f} + m\epsilon_{5d} + \frac{m(m-1)}{2} U^{GS}_{5d5d} \\ 
 & + 10mU^{GS}_{3d5d} + 10n U^{GS}_{3d4f} + mn U^{GS}_{4f5d} + k\varepsilon_{6s}  + \frac{k(k-1)}{2} U^{GS}_{6s6s} \\
 & +  mk U^{GS}_{5d6s} + nk U^{GS}_{4f6s} + 10k U^{GS}_{3d6s}
\end{split}
\end{equation}
\begin{equation}\label{M45nES} 
\begin{split}
  E_{ES}(n,m,k)  &= 9\varepsilon_{3d} + (n+1) \varepsilon_{4f} + \frac{n(n+1)}{2} U^{ES}_{4f4f} + m\varepsilon_{5d} + \frac{m(m-1)}{2} U^{ES}_{5d5d} \\ 
  & + 9mU^{ES}_{3d5d} + 9(n+1) U^{ES}_{3d4f} + (n+1)m U^{ES}_{4f5d} + k\varepsilon_{6s}  + \frac{k(k-1)}{2} U^{ES}_{6s6s} \\
 & +  mk U^{ES}_{5d6s} + (n+1)k U^{ES}_{4f6s} + 9k U^{ES}_{3d6s}
\end{split} 
\end{equation}
\end{subequations}
Where $n$, $m$, $k$ refer to the occupancy in the $4f$, $5d$, and $6s$ orbitals, respectively, while $\varepsilon_{\alpha}$ and $U_{\alpha, \beta}$ represent the on-site orbital energy and the inter- and intra-shell Coulomb integrals between $\alpha$ and $\beta$ orbitals, with $\alpha$, $\beta$ = $3d$, $4f$, $5d$, and $6s$. We define $E_{3d\rightarrow4f}$ as the energy difference given by equations~\ref{M45nES}~and~\ref{M45nGS}, representing the energy of the $M_{4,5}$ XAS transition.
\begin{equation}\label{deltaM45n} 
\begin{split}
E_{3d\rightarrow4f}(n,m,k)  &= \varepsilon_{4f} - \varepsilon_{3d} + n \left\{ \frac{n+1}{2} U^{ES}_{4f4f} - \frac{n-1}{2} U^{GS}_{4f4f} + 9U^{ES}_{3d4f} - 10U^{GS}_{3d4f} \right\} \\
& + m \left\{ \frac{m-1}{2} \left[ U^{ES}_{5d5d} - U^{GS}_{5d5d}\right] + 9 U^{ES}_{3d5d} - 10U^{GS}_{3d5d} + U^{ES}_{4f5d}\right\} \\ 
& + k \left\{ \frac{k-1}{2} \left[ U^{ES}_{6s6s} - U^{GS}_{6s6s}\right] + 9 U^{ES}_{3d6s} - 10U^{GS}_{3d6s} + U^{ES}_{4f6s}\right\} \\ 
& + nm (U^{ES}_{4f5d} - U^{GS}_{4f5d}) +  nk (U^{ES}_{4f6s} - U^{GS}_{4f6s}) + mk(U^{ES}_{5d6s} - U^{GS}_{5d6s}) + 9 U^{ES}_{3d4f}
\end{split}
\end{equation}
\subsection{1.2 $M_{2,3}$ XAS transition: $3p^{6}5d^{m}$ $\rightarrow$ $3p^{5}5d^{m+1}$} 
For this absorption edge, the ground state configuration is given by $3p^{6}4f^{n}5d^{m}6s^{k}$ configuration, while the excited state is given by $3p^{5}4f^{n}5d^{m+1}6s^{k}$. Following the same calculation as before, we obtain:
\begin{subequations}
\begin{equation}\label{M32nGS}
\begin{split}
 E_{GS}(n,m,k) &= 6\varepsilon_{3p} + n\varepsilon_{4f} + \frac{n(n-1)}{2} U^{GS}_{4f4f} + m\varepsilon_{5d} + \frac{m(m-1)}{2} U^{GS}_{5d5d} + k\varepsilon_{6s} + \frac{k(k-1)}{2} U^{GS}_{6s6s}  \\ 
 + &  6n U^{GS}_{3p4f} + 6mU^{GS}_{3p5d}  + 6k U^{GS}_{3p6s}+ nmU^{GS}_{4f5d}+ nk U^{GS}_{4f6s} + mk U^{GS}_{5d6s} 
\end{split}
\end{equation}
\begin{equation}\label{M32nES}
\begin{split}
  E_{ES}(n,m,k) &= 5\varepsilon_{3p} + n\varepsilon_{4f} + \frac{n(n-1)}{2} U^{ES}_{4f4f} + (m+1)\varepsilon_{5d} + \frac{m(m+1)}{2} U^{ES}_{5d5d} + k\varepsilon_{6s} + \frac{k(k-1)}{2} U^{ES}_{6s6s}  \\ 
  & + 5nU^{ES}_{3p4f} + 5(m+1)U^{ES}_{3p5d} + 
  5kU^{ES}_{3p6s} +
  nk U^{ES}_{4f6s} +
  n(m+1)U^{ES}_{4f5d} + (m+1)k U^{ES}_{5d6s} 
\end{split}
\end{equation}
\end{subequations}
We define $E_{3p\rightarrow5d}$ as the energy difference between the ground state and excited states given in equations~\ref{M32nGS}~and~\ref{M32nES}, respectively. This leads to the following formula:
\begin{equation}\label{deltaM32n}
\begin{split}
E_{3p\rightarrow5d}(n,m,k) &= \varepsilon_{5d} - \varepsilon_{3p} + n  \left\{ \frac{n-1}{2} \left[U^{ES}_{4f4f} - U^{GS}_{4f4f}\right] + 5U^{ES}_{3p4f} - 6U^{GS}_{3p4f} + U^{ES}_{4f5d} \right\} \\
& + m \left\{ \frac{m+1}{2} U^{ES}_{5d5d} -  \frac{m-1}{2} U^{GS}_{5d5d} + 5U^{ES}_{3p5d} - 6U^{GS}_{3p5d} \right\} \\
& + k \left\{ \frac{k-1}{2} \left[U^{ES}_{6s6s} -  U^{GS}_{6s6s}\right]  + 5U^{ES}_{3p6s} - 6U^{GS}_{3p6s} + U^{ES}_{5d6s} \right\} \\
& + nm  (U^{ES}_{4f5d} -  U^{GS}_{4f5d}) + 
nk (U^{ES}_{4f6s} -  U^{GS}_{4f6s}) + mk (U^{ES}_{5d6s} -  U^{GS}_{5d6s}) + 5U^{ES}_{3p5d}
\end{split}
\end{equation}

\subsection{1.3 Calculation of the energy shift induced by configurational changes from atoms to clusters}
For the sake of simplicity, we consider that all Coulomb integrals in the ground state and excited state are equal. This is justified by examining, for example, the dominant terms in Eq.~\eqref{deltaM45n}, namely $U_{3p4f}$ and $U_{4f4f}$, whose difference between the ground state and the excited state values is approximately 10\% at most. The same applies to the terms $U_{3d4f}$ and $U_{4f5d}$, which dominate in Eq.~\eqref{deltaM32n}. All this together leads to the generalized formula for both transitions:
\begin{equation}\label{Etrans}
\begin{split}
E_{c \rightarrow v}(n,m,k)  &= \varepsilon_{v} - \varepsilon_{c} + (4l_{c}+1) U_{cv} + \Vec{\textrm{q}} \cdot \overrightarrow{\Delta U}
\end{split}
\end{equation}
Here, $c$ and $v$ represent the core and valence (final) states, respectively, with $(c,v) = (3d,4f)$ for the $M_{4,5}$ edge and $(3p, 5d)$ for the $M_{2,3}$ edge. $l_{c}$ is the azimuthal quantum number of the core state which is equal to 1 or 2 for the $3p$ and $3d$ core states, and $U_{cv}$ is the Coulomb integral involving electrons in the core and final states. The vectors $\Vec{\textrm{q}}$ and $\overrightarrow{\Delta U}$ are defined as:
\[
\Vec{\textrm{q}} = \begin{bmatrix} n \\ m \\ k \end{bmatrix} \quad \text{and} \quad \overrightarrow{\Delta U} = \begin{bmatrix} U_{v,4f} - U_{c,4f} \\ U_{v,5d} - U_{c,5d} \\ U_{v,6s} - U_{c,6s} \end{bmatrix}
\]
Thus, we can calculate the difference in energy between the transitions of two different configurations defined by the triples ($n'$, $m'$, $k'$) and ($n$, $m$, $k$)  as follows: 
\begin{equation}\label{DeltaFinal}
\begin{split}
\Delta_{c\rightarrow v} &= E_{c \rightarrow v}(n',m',k') - E_{c \rightarrow v}(n,m,k)
\\
&= (4l_{c}+1)(U'_{cv} - U_{cv}) + (\Vec{\textrm{q}'} \cdot \overrightarrow{\Delta U'} - \Vec{\textrm{q}} \cdot \overrightarrow{\Delta U})   
\end{split}
\end{equation}
The equation~\ref{DeltaFinal} represents the general expression describing the energy differences measured in the XAS transitions from two electronic configurations of the same rare earth, dependent solely on multi-orbital Coulomb integrals. Additionally, by exploiting the approximation of the linear response of $U$ to the occupancy of different orbitals~\cite{cococioni_linear_2005}, we can extend the applicability of Eq.~\eqref{DeltaFinal} to include fractional occupancies of the orbitals. 
For the sake of generality, we define $\textrm{q}_{4f}$, $\textrm{q}_{5d}$ and $\textrm{q}_{6s}$ ($\in\mathbb{R}$) as the non-integer occupations of the $4f$, $5d$ and $6s$ orbitals. 

This approximation allows us to utilize Eq.~\eqref{DeltaFinal} to address fractional configurational changes due to cluster nucleation observed in our measurements, which involve partial charge transfer between orbitals. To this extent, we replace $U' = U^{\textrm{cl}}$ and $U = U^{\textrm{at}}$ in Eq.~\eqref{DeltaFinal}, both 'cl' and 'at' indexes refer to clusters and atoms, respectively. The simplest approach for Eq.~\eqref{DeltaFinal} is a mean-field (MF) approximation, where we consider that  $U^{\textrm{at}} = U^{\textrm{cl}}$. This approximation is justified by the fact that the difference in $U$ parameters induced by a change in the $4f$ occupation, namely, $4f^{n} \rightarrow 4f^{n-1}$ transition, is at maximum 5\%. From this, we get the general formula: 
\begin{equation}\label{DeltaMFGen}
\Delta_{c\rightarrow v}^{\textrm{MF}} = \overrightarrow{\Delta \textrm{q}} \cdot \overrightarrow{\Delta U} =
\sum_{\alpha} \delta \textrm{q}_{\alpha} (U_{v,\alpha} - U_{c,\alpha}), 
\end{equation}
where $\alpha$ labels the $4f, 5d, 6s$ states considered in this work. We note that Eq.~\eqref{DeltaMFGen} represents the most general formulation involving a multi-orbital description and can be extended to any orbital $\alpha$ (e.g. $6p$, \textrm{etc.}) for any XAS transitions (e.g. $M_{1}$ edge~\cite{singha_mapping_2021}). For lanthanide atoms, mapping electronic and magnetic properties with orbital resolution involves measuring the $M$ edges, specifically $M_{4,5}$, $M_{2,3}$, and $M_{1}$, corresponding to transitions $3d \rightarrow 4f$, $3p \rightarrow 5d6s$, and $3s \rightarrow 6p$, respectively~\cite{singha_mapping_2021}. In our study, we measured the $M_{4,5}$ and $M_{2,3}$ edges, which involve $4f$, $5d$, and $6s$ orbitals. The explicit expressions derived from Eq.~\eqref{DeltaMFGen} are:
\begin{equation}\label{DeltaMF}
 \Delta_{c\rightarrow v}^{\textrm{MF}} =
 \begin{bmatrix} \textrm{q}_{4f}^{\textrm{cl}}-\textrm{q}_{4f}^{\textrm{at}} \\ \textrm{q}_{5d}^{\textrm{cl}}-\textrm{q}_{5d}^{\textrm{at}} \\ \textrm{q}_{6s}^{\textrm{cl}}-\textrm{q}_{6s}^{\textrm{at}} \end{bmatrix} 
\cdot  
\begin{bmatrix} 
U_{v,4f} - U_{c,4f} \\
U_{v,5d} - U_{c,5d} \\ 
U_{v,6s} - U_{c,6s}
\end{bmatrix}
\end{equation}
Note that the exact expression for the Coulomb integrals is given by the matrix elements of the electron-electron interaction, which  is considered in the non-relativistic limit (no spin-orbit coupling) and, therefore, in the base $\ket{\alpha} = \ket{\psi_{n, l , m_{l}, s, m_{s}}}$ where $n, l, m_{l}, s, m_{s}$ are the quantum numbers discriminating each shell. The multipole development of the matrix elements in the basis of spherical harmonics can be written as:
\begin{equation}\label{U}
    \begin{split}
    U_{\alpha, \beta} &= \bra{\alpha}\frac{e^{2}}{|r_{\alpha} - r_{\beta}|}\ket{\beta} \\ &= \underbrace{\sum_{k}^{2l}\sum_{\alpha}^{shells} f_{k}(l_{\alpha},l_{\alpha})F^{k}(l_{\alpha},l_{\alpha})}_{\text{Direct exchange (equiv.)}} +\underbrace{\sum_{k}^{2l}\sum_{\alpha,\beta,\alpha\neq \beta}^{shells}f_{k}(l_{\alpha},l_{\beta})F^{k}(l_{\alpha},l_{\beta})}_{\text{Indirect exchange (equiv.)}} 
    \\ &+ \underbrace{\sum_{k}^{2l}\sum_{\alpha,\beta,\alpha\neq \beta}^{shells}g_{k}(l_{\alpha},l_{\beta})G^{k}(l_{\alpha},l_{\beta})}_{\text{Indirect exchange (inequiv.)}}
\end{split}
\end{equation}
From Eq.~\eqref{U}, we can see that the multi-orbital Coulomb interactions are provided by a linear combination of Slater integrals involving direct and indirect exchange interaction terms between equivalent and inequivalent electrons from the same and/or different shells. From this, a reasonable approximation can be made and which consists of considering that $U_{\alpha, \beta} \propto F^{0}_{\alpha, \beta}$, where $F^{0}_{cv}$ denotes the first-order term of the Slater integral obtained from the Cowan code~\cite{cowan_theory_nodate}. This code utilizes the multiconfiguration Hartree-Fock method, and we have averaged the values for all $4f(5d6s)$ configurations, resulting in effective $U$ values. Moreover, and as mentioned in the main text, the substrate's role is to screen the Coulomb integral values, necessitating the introduction of an orbital-dependant screening parameter $\kappa_{\alpha}$. This parameter reflects the rescaling of electron-electron interaction due to the substrate's electrons, ultimately renormalizing the Coulomb integrals as $U_{\alpha, \beta} = \sqrt{\kappa_{\alpha}\kappa_{\beta}}F^{0}_{\alpha,\beta}$. The screening parameters $\kappa_{\alpha}$ are obtained as part of the fitting procedure described in the next section, and are used to obtain the value for all $U_{\alpha, \beta}$ from the computed values of $F^{0}_{\alpha,\beta}$.

During cluster nucleation, the change in occupancy induced in the $4f$, $5d$ and $6s$ orbitals is defined as $\delta\textrm{q}_{4f} = \textrm{q}_{4f}^{\textrm{cl}} - \textrm{q}_{4f}^{\textrm{at}}(= 0~\textrm{or}~-1)$, $\delta\textrm{q}_{5d}=\textrm{q}_{5d}^{\textrm{cl}} - \textrm{q}_{5d}^{\textrm{at}}$ and $\delta\textrm{q}_{6s}=\textrm{q}_{6s}^{\textrm{cl}} - \textrm{q}_{6s}^{\textrm{at}}$ respectively. Following the main text, we define $\Delta_{5} \equiv \Delta_{3d\rightarrow 4f}^{\textrm{MF}}$ and $\Delta_{3} \equiv \Delta_{3p\rightarrow 5d}^{\textrm{MF}}$ as the energy difference for the XAS maximima, between clusters and atoms, for the $M_{4,5}$ and $M_{2,3}$ edges, respectively. This leads to the following expressions:
\begin{subequations}
\begin{equation}
\label{eq:M45}
\begin{split}
\Delta_{5} &=  
\delta\textrm{q}_{4f}(\underbrace{\kappa_{4f}F^{0}_{4f4f} - \sqrt{\kappa_{3d}\kappa_{4f}} F^{0}_{3d4f})}_{U_{4f4f} - U_{3d4f}} + \delta\textrm{q}_{5d}(\underbrace{\sqrt{\kappa_{4f}\kappa_{5d}}F^{0}_{4f5d} - \sqrt{\kappa_{3d}\kappa_{5d}} F^{0}_{3d5d})}_{U_{4f5d} - U_{3d5d}} \\
&\quad + \delta\textrm{q}_{6s}(\underbrace{\sqrt{\kappa_{4f}\kappa_{6s}}F^{0}_{4f6s} - \sqrt{\kappa_{3d}\kappa_{6s}} F^{0}_{3d6s})}_{U_{4f6s} - U_{3d6s}}
\end{split}
\end{equation}
\begin{equation}
\label{eq:M23}
\begin{split}
\Delta_{3} &= \delta\textrm{q}_{4f}(\underbrace{\sqrt{\kappa_{4f}\kappa_{5d}} F^{0}_{4f5d} - \sqrt{\kappa_{3p}\kappa_{5d}} F^{0}_{3p5d})}_{U_{4f5d} - U_{3p5d}} + \delta\textrm{q}_{5d} (\underbrace{\kappa_{5d}F^{0}_{5d5d} - \sqrt{\kappa_{3p}\kappa_{5d}} F^{0}_{3p5d})}_{U_{5d5d} - U_{3p5d}} \\
&\quad + \delta\textrm{q}_{6s} (\underbrace{\sqrt{\kappa_{5d}\kappa_{6s}} F^{0}_{5d6s} - \sqrt{\kappa_{3p}\kappa_{6s}} F^{0}_{3p6s})}_{U_{5d6s} - U_{3p6s}}
\end{split}
\end{equation}
\end{subequations}

\subsection{1.4 Bayesian optimization to extract $\delta \textrm{q}^{\alpha}$, $\kappa_{\alpha}$ and $U_{\alpha, \beta}$}
\begin{figure}[h!]
        \centering \includegraphics[width=0.7\textwidth]{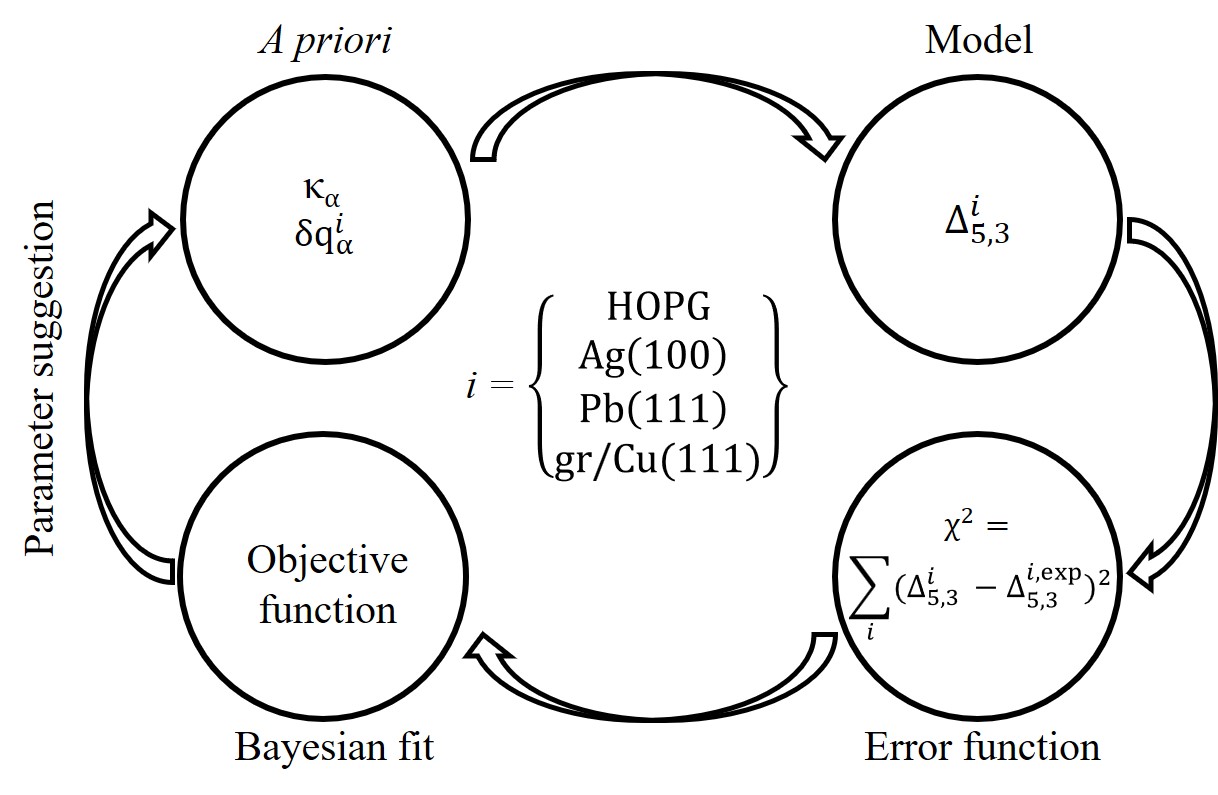}
        \caption{Flow-chart of the Bayesian optimization combined with equations.~\eqref{eq:M45} and ~\eqref{eq:M23} used to fit the experimental values of $\Delta_{5,3}$.}
        \label{fig:SI0}
\end{figure}

To determine the values of $\kappa_{\alpha}$ and $\delta\textrm{q}_{\alpha}$, we employed a supervised machine learning program based on a Bayesian optimization (BO) algorithm~\cite{frazier_tutorial_2018} in MATLAB. This program aimed to match the experimental values of $\Delta_{5,3}$ (see Fig.~\ref{fig:SI0}). The assumption of identical renormalization for the core and $4f$ states is justified by their limited impact from hybridization with the substrate, while we renormalize each outer shell with a different $\kappa_{\alpha}$. This approach helps to avoid overparametrization of the problem. Furthermore, the use of BO optimization approach is justified by the fact that the analytical solutions of these equations lie outside the physical boundaries of number of electrons  within each orbital. Therefore, we aim to find the set of parameters that come as close as possible to the experimental values. To do so, the $\kappa_{\alpha}$ and $\delta \textrm{q}_{\alpha}$ parameters were determined by iterative minimizing the error function, specifically the residual function ($\chi^{2}$), through fitting the 8 independent energy shifts extracted from the $M_{5,3}$ XAS spectra on each substrate. To minimize the number of free parameters in the model, we make the assumption that the orbitals that are not significantly involved in bond formation (i.e., $3p, 3d,4f$) will be renormalized by the same parameter, $\kappa_{3p} = \kappa_{3d} = \kappa_{4f}$. 

The optimization is achieved by using a prior distribution as input for an acquisition function, such as the lower confidence bound function we employed, that guides the selection of the next data point in the parameter space. Once the objective function is evaluated at this selected point, the sampled data is used to fit a Gaussian process regression. We utilized the \textit{fitrgp} function for this purpose, resulting in a posterior distribution characterized by a mean and a standard deviation. Importantly, we do not specify an initial prior. Instead, we randomly sample a certain number of points across the parameter space before the regression process (in our case, just one point). This posterior distribution then serves as the prior for the next iteration. Concretly, we initiated the optimization with random parameters and use Eqs.~\eqref{eq:M45} and~\eqref{eq:M23} to fit the various experimental values of $\Delta_{5}$ and $\Delta_{3}$ for each substrate, which are the input parameters. The calculated values were then compared with the experimental ones to compute the $\chi^{2}$, and BO was used to minimize this error function. In this work, we check the convergence of  the procedure by varying the number of steps from 200 to 1000, which results in very similar output parameters. 

\noindent From Eqs.~\eqref{eq:M45}-\eqref{eq:M45}, we will discuss two cases: a) the two-orbital model, i.e. take into account the 
$4f$ and $5d$ orbitals; b) the three-orbital model, i.e. including also the $6s$ orbital.

\subsubsection{a) The two-orbital model:}
In the two-orbital model, we consider the $4f$ and $5d$ orbitals, and the last two terms in Eqs.~\eqref{eq:M45}-\eqref{eq:M23} are omitted. In this approach, the environment consists of the $6s$ orbital and the substrate.
For the BO, the number of parameters is 6, namely, 2 global parameters ($\kappa_{3p,3d,4f}$ and $\kappa_{5d}$), as well as 4 parameters ($\delta q_{5d}$) related to the $5d$ orbitals, one for each substrate. The fit results for $\kappa_{\alpha}$ and $U_{\alpha, \beta}$ are presented in Tab.~\ref{tab:comparisonwo6s} ($\delta q_{5d}$ are presented in the main text). 
\begin{table*}[htbp]
\caption{Comparison between the Coulomb potential for the bare Nd atom and the rescaled Coulomb potential for Nd atoms supported on the surface, \underline{in absence of $6s$ orbitals}. All units are in eV. The rescaling factors on each orbital are shown in the two first columns.}
\label{tab:comparisonwo6s}
\begin{ruledtabular}
\begin{tabular}{@{\hspace{2mm}} cc  || @{\hspace{1mm}}  p{2.5cm} *{8}{c}}
$\alpha$ & $\kappa_{\alpha}$ & System & $U_{4f4f}$ & $U_{3d4f}$ & $U_{4f5d}$ & $U_{3d5d}$ &  $U_{3p4f}$ & $U_{3p5d}$ & $U_{5d5d}$ & \\
\midrule
$3p,3d,4f$ & 0.29 & Bare Nd & 27.3 & 37.4 & 12.1 & 12.9 & 37.3 & 12.9 & 9.6   \\
 $5d$ & 0.06 & Screened values & 7.8 & 10.8 & 1.6 & 1.7  & 10.7 & 1.7 & 0.6  \\
\end{tabular}
\end{ruledtabular}
\end{table*}
Comparing with the literature, the extracted screening values closely align with those found for cerium fluorosulfide pigments~\cite{tomczak_rare-earth_2013, aryasetiawan_electronic_nodate}, early rare-earth sesquioxides~\cite{tomczak_spectral_2010}, and with the empirical model developed by Van der Marel and Sawatzky~\cite{van_der_marel_electron-electron_1988}, for various $f$ shell occupancy in RE metals ($0.19\leq\kappa\leq0.28$), based on X-ray photoemission spectroscopy and bremsstrahlung isochromat spectroscopy data~\cite{lang_study_1981, cox_study_1981}. Furthermore, the rescaling parameters $\kappa_{\alpha}$ decrease with the increasing spatial extension of the orbitals owing to the increase of electron-electron screening~\cite{van_der_marel_electron-electron_1988}. This last observation allows us to justify the choice of considering the same screening for the very localized $3p$, $3d$ and $4f$ orbitals. Note that the proposed method, which aims to measure intra-atomic Coulomb integrals for REs with surface sensitivity, bridges the gap between experimental results and various theoretical approaches, including density functional theory calculations~\cite{kozub_electronic_2016, donati_correlation_2021, singha_mapping_2021}, Anderson impurity models~\cite{grazioli_characteristic_2001, gunnarsson_electron_1983}, and \textit{ab-initio} and dynamical mean field calculations~\cite{tomczak_spectral_2010, tomczak_rare-earth_2013, aryasetiawan_electronic_nodate}.

\subsubsection{b) The three-orbital model:}\label{w6s}
In this approach, $4f$, $5d$, and $6s$ orbitals are taken into account, and considering the substrate as the environment. This results in 11 parameters for the BO process, comprising 3 global values for the screening ($\kappa_{3p,3d,4f}$ and $\kappa_{5d}$, as in the two-orbital model, plus $\kappa_{6s}$) and substrate-dependant sets of $\delta\textrm{q}_{\alpha}$ parameters (2 for each substrate: 1 for $5d$ and 1 for $6s$).
Through this process, we extracted all the $\delta\textrm{q}_{\alpha}$, $\kappa_{\alpha}$ and using these later parameters, the entire set of $U_{\alpha,\beta}$ can be computed, as listed in Tab.~\ref{tab:comparisonw6s}.

\begin{table}[htbp]
    \centering
    \caption{Top: Charge variation for Nd on different substrates obtained from the fit to the data using Eqs.~\eqref{eq:M45} and \eqref{eq:M23}, on different substrates. The rescaling factors on each orbital are shown in the two last columns. Bottom: Comparison between the Coulomb potential for the bare Nd atom and the rescaled Coulomb potential for Nd atoms supported on the surface, \underline{in presence of $6s$ orbitals}. The values presented are rounded to the first digit and the units are in eV.}
    \label{tab:comparisonw6s}
    \begin{tabular}{@{}c@{}}
        \begin{minipage}[t]{0.6\textwidth}
            \centering
                \begin{tabular}{lcccc || @{\hspace{2mm}} cc }
                    \toprule
                     \toprule
                    Nd on & {HOPG} & {Gr/Cu(111)} & {Ag(100)} & {Pb(111)} & $\alpha$ & $\kappa_{\alpha}$ \\
                    \midrule
                    $\delta \textrm{q}_{4f}$ & -1 & -1 & -1 & 0 & $3p,3d,4f$ & 0.29\\
                    $\delta \textrm{q}_{5d}$ & 0.39 & 0.13 & 1.14 & 0.42 & $5d$ & 0.07\\
                    $\delta \textrm{q}_{6s}$ & 0.84 & 0.99 & 0.13 & 0.05 & $6s$ & 0.0004 \\
                \end{tabular}
        \end{minipage} 
        \\
        \begin{minipage}[t]{1.0\textwidth}
                \begin{tabular}{ @{\hspace{1mm}}  p{2.5cm} *{11}{c}}
                    \toprule
                    System & $U_{4f4f}$ & $U_{3d4f}$ & $U_{4f5d}$ & $U_{3d5d}$ & $U_{4f6s}$ & $U_{3d6s}$ & $U_{3p4f}$ & $U_{3p5d}$ & $U_{5d5d}$ & $U_{3p6s}$ & $U_{5d6s}$ \\
                    \midrule
                     Bare Nd & 27.3 & 37.4 & 12.1 & 12.9 & 7.6 & 7.9 & 37.3 & 12.9 & 9.6 & 7.8 & 6.7  \\
                    Screened values & 7.9 & 10.8 & 1.7 & 1.8 & 0.1 & 0.1 & 10.8 & 1.8 & 0.6 & 0.1 & 0.0  \\
                    \bottomrule
                    \bottomrule
                \end{tabular}
        \end{minipage}
    \end{tabular}
\end{table}
On all substrates, the changes in occupancy of the $5d$ orbitals are very similar (around $\approx 0.1\textrm{e}^{-}$) to the scenario where the contribution of the $6s$ orbitals is omitted, as detailed in Tab.~1, in the main text. Additionally, a preference for filling $6s$ orbitals is evident in carbonated substrates~\cite{pivetta_measuring_2020, curcella_valence_2023}.
Moreover, when comparing the screened Coulomb integrals obtained in Tab.~\ref{tab:comparisonw6s} with those reported in Tab.~\ref{tab:comparisonwo6s}, we observe that the fit results for the two-orbital model are robust. This is evidenced by the energy differences of the last terms in Eqs.~\eqref{eq:M45}-\eqref{eq:M23} being close to $\approx 0$~meV ($U_{4f6s} - U_{3d6s}$) and $\approx 100$~meV ($U_{5d6s} - U_{3p6s}$), for a maximum change of $\delta \textrm{q}_{6s}$ ($\approx 1\textrm{e}^{-}$) on Gr/Cu(111).
Probing these differences necessitates significantly higher energy resolution due to the shorter lifetime of the $3p$ core hole compared to the $3d$ state. Therefore, the approximation of discarding the $6s$ orbitals, as outlined in the main text, is justified.

\section{2. Additional XAS and XMCD data}    
\subsection{2.1 XAS/XMCD at the $M_{4,5}$ edges}

In this section, we present all the XAS/XMCD spectra at $M_{4,5}$ edges, at normal incidence (NI) and grazing incidence (GI, $60^\circ$ off normal) at $6\, \text{T}$ and $1.8\, \text{K}$, for all the coverages studied on HOPG, Ag(100), and Pb(111).

\subsubsection{2.1.1 Calibration of Nd coverage}
To calibrate the coverage of Nd atoms on different substrates, we used a reference calibration with holmium (Ho) single atoms on Ag(100)~\cite{donati_magnetic_2016}. For Ho on Ag(100), 1\% of a monolayer (ML) corresponds to a  an integral of the XAS whiteline $t_{\mathrm{Ho/Ag(100)}}^{\mathrm{1\%~ML}}~=~0.17$. 
Then, we rescale this calibration XAS integral of Nd on Ag(100) by considering the ratio of the number of holes between Nd and Ho. For the Nd, we considered a value $n_{h}^{\textrm{Nd}}$~=~10.5 as an intermediate configuration between atomic-like and bulk-like configurations. Thus, for Nd, we obtained the following calibration:  $t_{\mathrm{Nd/Ag(100)} 
}^{\mathrm{1\%~ML}} = \cfrac{n_{h}^{\textrm{Nd}}}{n_{h}^{\textrm{Ho}}} \times t_{\mathrm{Ho/Ag(100)} 
}^{\mathrm{1\%~ML}}=0.45$.

On the other substrates, we should take into account two additional corrections. First, we need to consider the difference in electronic densities on the substrate which leads to different background contributions in the total electron yield (TEY). Using the X-ray database of the Center for X-ray Optics (CXRO) at the Lawrence Berkeley Laboratory~\cite{HENKE1993181}, we calculate the absorption of 10~nm thick carbon, lead and copper at 965~eV to estimate the relative TEY. Second, we considered the difference in adsorption site density, which is is determined by the ratio of the number of adsorbing sites and the unit cell area. All this together leads to the following formula for the coverage ($\Theta$~(\%~ML)):
\begin{equation}
\Theta~(\%~\mathrm{ML}) = t_{\mathrm{Nd/substrate}}^{\mathrm{measured}} \times \underbrace{\cfrac{\text{\small background of the substrate}}{\text{\small background of Ag(100)} \times t_{\mathrm{Nd/Ag(100)}}^{\mathrm{1\%~ML}}}}_{\text{\small Background correction}} \times \underbrace{\cfrac{\text{\small density of adsorbing sites on Ag(100)}}{\text{\small density of adsorbing sites on the substrate}}}_{\text{\small Adsorption site density correction}}
\end{equation}

This permits to calculate all the coverages, for all HOPG, Ag(100) and Pb(111), as it is shown in Tab.~\ref{tab:coverage}. However, in the case of graphene on copper substrate (Gr/Cu(111)), determining the surface absorption using CXRO~\cite{HENKE1993181} is not accurate, since the substrate consists of two different materials with very different absorptions. For this reason, another strategy has been employed, which consists of using the pre-edge ratio between Gr/Cu(111) and HOPG ($\approx 6.16$), provided that the TEY and normalization current ($I_{0}$) electrometer ranges are the same, which is the case here. In addition, since Gr/Cu(111) and HOPG have the same elementary lattice structure, their adsorption site densities are the same. Thus, $\Theta_{\mathrm{Nd/Gr/Cu(111)}}~(\%~\mathrm{ML})~=~\underbrace{\cfrac{\textrm{Pre-edge Gr/Cu(111) at 965~eV}}{\textrm{Pre-edge HOPG at 965~eV}}}_{\textrm{Correction with the pre-edge}} \times \cfrac{t_{\mathrm{Nd/Gr/Cu(111)}}^{\mathrm{measured}}}{t_{\mathrm{Nd/HOPG}}^{\mathrm{1\% ML}}}$, and the corresponding coverages are shown in the last two lines of Tab.~\ref{tab:coverage}.
\begin{table}[h]
\caption{All measured $t$ integrals on the different substrates presented in the paper translated in terms of Nd coverage ($\Theta$~(\%~ML)). The $t$ values for 1\%~ML of Nd on Ag(100), HOPG, Pb(111) and on Gr/Cu(111) are, respectively, 0.45, 5.5, 0.35 and 0.89.}
\label{tab:coverage}
\centering
\begin{tabular}{|c|c|c|c|c|}
\hline
\textbf{Substrate} & \textbf{$t_{\mathrm{Nd/substrate}}^{\mathrm{measured}}$} & \textbf{Substrate absorption at 965~eV} & \textbf{Adsorption site density (nm$^{-2}$)} & \textbf{$\Theta$~(\%~ML)} \\ \hline
\multirow{3}{*}{Ag(100)} & 0.24 & \multirow{3}{*}{0.06}  & \multirow{3}{*}{11.96} & 0.54 \\ \cline{2-2}\cline{5-5}
 & 0.50  & & & 1.12 \\ \cline{2-2}\cline{5-5}
 & 2.16 & & & 4.84 \\ \hline
\multirow{3}{*}{HOPG} & 2.28 & \multirow{3}{*}{0.0043} & \multirow{3}{*}{9.68} & 0.41 \\ \cline{2-2}\cline{5-5}
 & 4.66 &  & & 0.85 \\ \cline{2-2}\cline{5-5}
 & 24.23 & & & 4.41 \\ \hline
\multirow{3}{*}{Pb(111)} & 0.27 & \multirow{3}{*}{0.06} & 
\multirow{3}{*}{8.16} & 
0.77 \\ \cline{2-2}\cline{5-5}
 & 0.51 &  & & 1.46 \\ \cline{2-2}\cline{5-5}
 & 2.35 & & & 6.74 \\ \hline
\multirow{2}{*}{Gr/Cu(111)} & 0.70 & \multirow{2}{*}{---} & \multirow{2}{*}{9.68}  & 0.78 \\ \cline{2-2}\cline{5-5}
 & 3.93 & & & 4.40 \\ \hline
\end{tabular}
\end{table}

\subsubsection{2.1.2 Sum rules on the XAS/XMCD at the $M_{4,5}$ edges}

\begin{figure}[h!]
        \centering \includegraphics[width=1.0\textwidth]{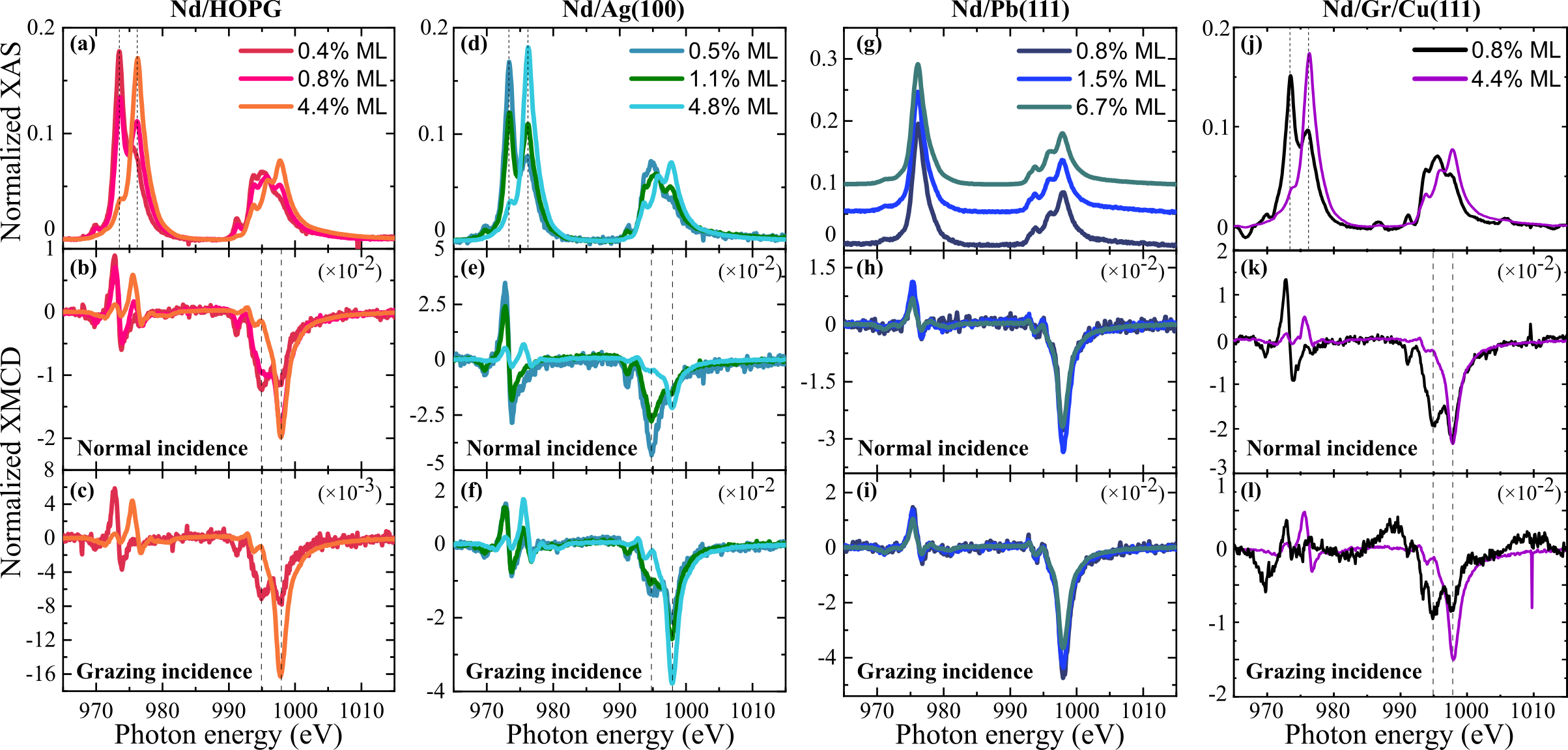}
        \caption{XAS and XMCD spectra in normal and grazing incidences on HOPG (a, b, c), Ag(100) (d, e, f), Pb(111) (g, h, i) and Gr/Cu(111) (j, k, l) respectively, at the $M_{4,5}$ edge of Nd.
        All spectra are normalized to the integral of the $M_{4,5}$ edge. Dashed lines mark the XAS and XMCD maxima for each $4f$ configuration. The Nd coverage in \%~ML is indicated in the legend.}
        \label{fig:SI1}
\end{figure}

As discussed in the main text on HOPG, Ag(100) and Gr/Cu(111), and as shown in Figs.~\ref{fig:SI1}a/d, we observe the modification of the XAS whiteline and the growth of the $4f^{3}$ peak during cluster nucleation, which finally leads to an energy shift of the whole spectrum. This shift is measured through the shift of the XAS maximum ($\Delta_{5} = 2.9\, \text{eV}$) when the whole investigated system is essentially composed of clusters. The same observation can be made on the XMCD spectra in Figs.~\ref{fig:SI1}b/c/e/f/k/l (independent of the angle of incidence), with an energy shift ($\Delta_{4} = \Delta_{5}$) observed at the $M_{4}$ edge of the XMCD. The angular dependence of the XMCD spectra of Nd on HOPG and Gr/Cu(111) (resp. Nd/Pb(111)) shows that the anisotropy easy-axis is out-of-plane (resp. in-plane) for all coverages, whereas on Ag(100), a reorientation from out-of-plane to in-plane is observed~\cite{singha_magnetic_2016}. 

The quantification of the orbital and spin moments for $4f$ electrons can be calculated using the following sum rules~\cite{thole_x-ray_1992, carra_x-ray_1993}:
\begin{subequations}
\begin{equation}\label{eq:4forb}
\mu_{L}^{4f} = -\cfrac{\mu_{B}}{\hbar} \langle \hat{L}_z^{4f} \rangle =  -n_{h}^{4f} \cfrac{2q}{t}
\end{equation}
\begin{equation}\label{eq:4fspin}
\mu_{S+D}^{4f}  = -\cfrac{\mu_{B}}{\hbar} \left( 2 \langle \hat{S}_z^{4f} \rangle + 6 \langle \hat{T}_z^{4f} \rangle \right) = -n_{h}^{4f} \cfrac{5p-3q}{t}
\end{equation}
\end{subequations}
With $t$ the integral of the XAS, $p$ the integral of the XMCD signal over $M_{5}$ edge, $q$ the integral of the XMCD signal over $M_{5}$ + $M_{4}$ edges, and $n_{h}^{4f}$ the number of the $4f$ holes in the ground state. Note that the sum rules do not provide direct access to the spin moment due to the contribution of the dipolar term \(D = 6 \langle \hat{T}_z \rangle\), where \(\langle \hat{T}_z \rangle\) is the expectation value of the magnetic dipole operator. The expectation value of the \(\langle \hat{T}_z \rangle\) operator in the $4f^{4}$ and $4f^{3}$ configurations have been obtained from the multiplet calculation of the free atom and are, respectively, -0.07~$\mu_{B}$ and 0.35~$\mu_{B}$ in normal incidence, and half the value at 60$^\circ$. To isolate the contribution of the spin moment ($\mu_{S}$), we can use the following expression:
\[
\mu_{S+D} = -  \underbrace{\cfrac{2\mu_{B}}{\hbar} \langle \hat{S}_z}_{\mu_S} \rangle \left(1+\frac{3\langle \hat{T}_z \rangle}{\langle \hat{S}_z \rangle}\right)
\]
At the lowest coverage, one can consider that one has a collection of single atoms, so a pure $4f^{4}$ (resp. $4f^{3}$) XMCD signal for Nd on HOPG, Ag(100) and Gr/Cu(111) (resp. $4f^{3}$). However, in the case of  
intermediate coverages, the XMCD signal is a mixture between single atoms and clusters, which in principle can be resolved by deconvolution of XAS~\cite{mocuta_artificial_2021} into pure $4f^{4}$ and $4f^{3}$ contributions, what we are unable to do. In addition, for the $4f^{3}$ configuration, different cluster sizes coexist, such as dimers, trimers, etc.~\cite{brune_microscopic_1998}. From the XAS and XMCD spectra, it is only possible to access the sum of $p$, $q$ and $t$ values for each species, leading to an estimation of the magnetic moments. Furthermore, in the case of a change in the electronic configuration from $4f^{4}$ to $4f^{3}$, the number of holes varies from $10$ to $11$, and thus we approximated that $n_{h} = 10.5$ (error of $5\%$).
The observed trend is not affected by these limitations.

\begin{figure}[h!]
 \centering \includegraphics[width=\textwidth]{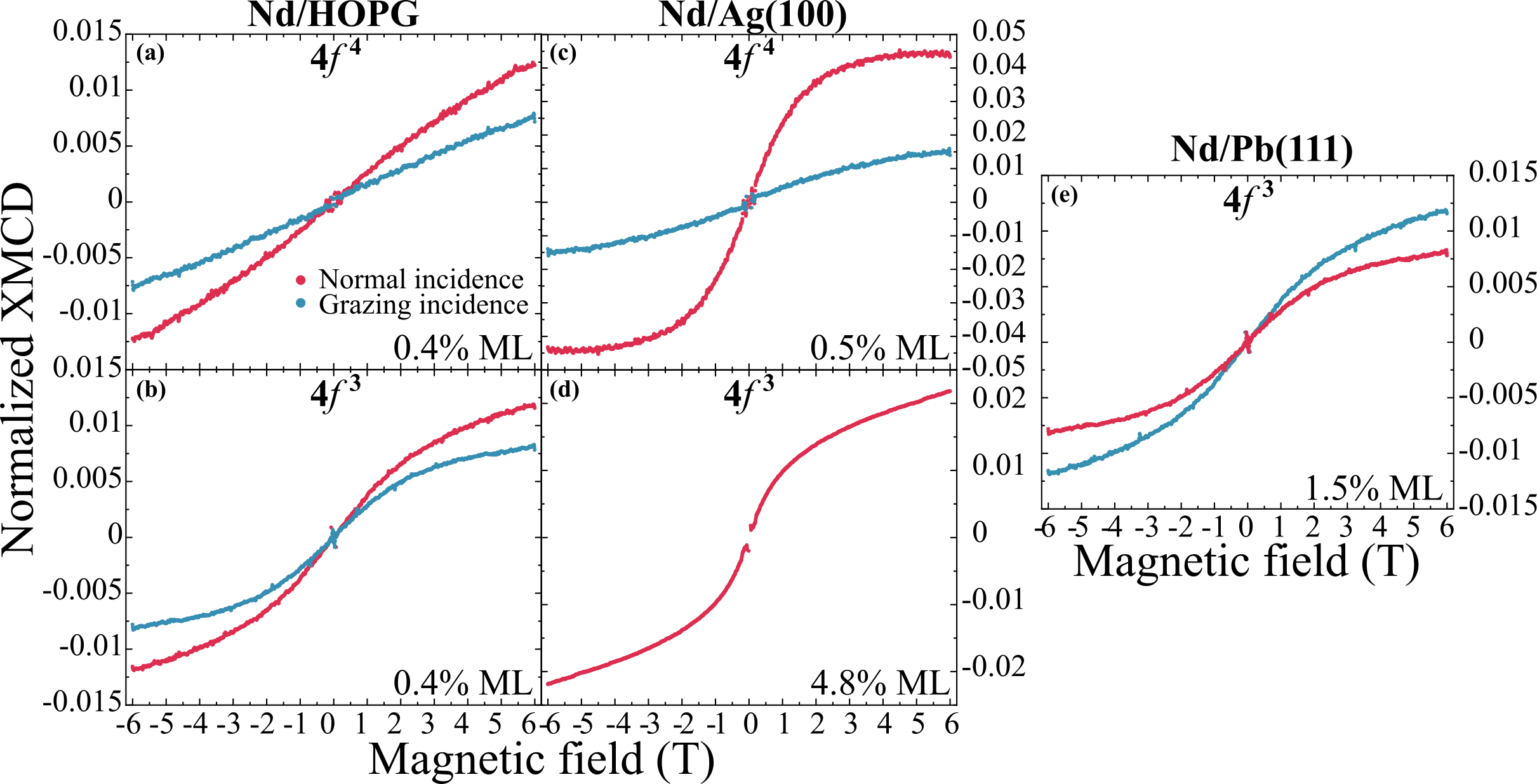}
  \caption{Variation of the XMCD signal at $M_{4}$ edge of Nd as a function of magnetic field for the $4f^{4}$ and $4f^{3}$ electronic configurations on HOPG (a, b) and on Ag(100) (c, d), respectively, and on Pb(111) (d, $4f^{3}$ configuration only). Red and blue curves represent the normal and grazing incidences, respectively. The Nd coverage in \%~ML is indicated in the legend.}
   \label{fig:SI2}
\end{figure}

To investigate the different contributions leading to the decrease in the orbital and spin moment values as extracted from sum rules, we measured the magnetization curves of Nd on the different substrates, as shown in Fig.~\ref{fig:SI2}. They were obtained from the magnetic field dependence of the maximum of the XMCD at the $M_{4}$ edge, for the atomic and bulk configurations. To simplify the discussion, we consider  different scenarios by substrate:
\begin{itemize}
    \item \textbf{On HOPG:} Fig.~\ref{fig:SI2}a shows the magnetization curve for Nd in its $4f^{4}$ configuration and exhibits a linear shape, suggesting the existence of antiferromagnetic interactions between the atoms on this surface. For the clusters, the magnetization curve becomes \textit{S-like} but does not saturate at 6~T, as shown in Fig.~\ref{fig:SI2}b. In both configurations, the magnetic easy-axis is out-of-plane.
    \item \textbf{On Ag(100):} Nd atoms in the $4f^{4}$ configuration, the magnetization in normal incidence reaches saturation above 4~T, as shown in Fig.~\ref{fig:SI2}c. Moreover, comparison with the grazing incidence curve indicates a strong out-of-plane anisotropy. In the bulk-like configuration shown in Fig.~\ref{fig:SI2}d, it can be seen that the magnetization curve does not saturate, and there is a tendency for the magnetization to increase as the magnetic field increases, which is characteristic of the existence of non-collinear magnetic states~\cite{van_dijk_unusual_2015}. The easy-axis of magnetization switches from out-of-plane to in-plane when clusters are formed, as it was also shown for other lanthanides on metals~\cite{singha_magnetic_2016}.
    \item \textbf{On Pb(111):} Due to the strong background of Pb(111), we have only measured the magnetization curve of the $4f^{3}$ population for the 1.5\%~ML coverage. The magnetization curve does not saturate due to magnetic anisotropy, with an easy-axis that is out-of-plane, for all coverages.
\end{itemize}
Through all these diverse scenarios, we conclude that the tendency of the decrease of $\mu_{L}$ and $\mu_{S}$ upon clustering can be ascribed to both the modification of the magnetic anisotropy~\cite{singha_magnetic_2016} between atoms and clusters, and the emergence of non-collinear magnetic states~\cite{van_dijk_unusual_2015} within the cluster.

\clearpage
\subsection{2.2 XAS/XMCD at the $M_{2,3}$ edges}
\begin{figure}[h!]
        \centering \includegraphics[width=\textwidth]{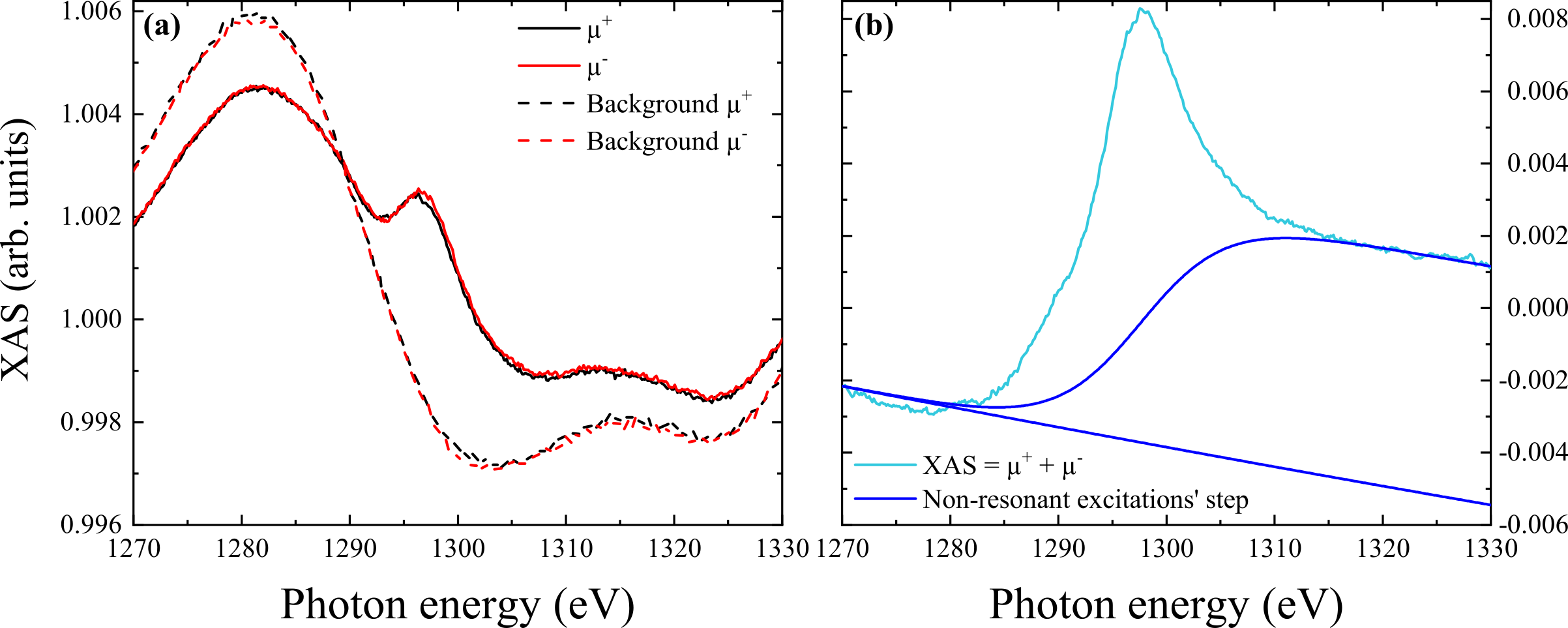}
        \caption{(a) Circular right ($\mu^{+}$ in black) and circular left ($\mu^{-}$ in red) $M_{3}$ XAS (raw data in full lines) for Nd on Ag(100), with their respective backgrounds (dotted lines) taken on pristine Ag(100). (b) $M_{3}$ XAS (cyan) and the corresponding non-resonant excitation background (blue), with their difference shown in Fig.\ref{fig:SI5}c. These data are taken for $\Theta$~=~4.8\%~ML, B~=~6~T, and T~=~1.8~K.}
        \label{fig:SIDT}
\end{figure}

\begin{figure}[h!]
        \centering \includegraphics[width=\textwidth]{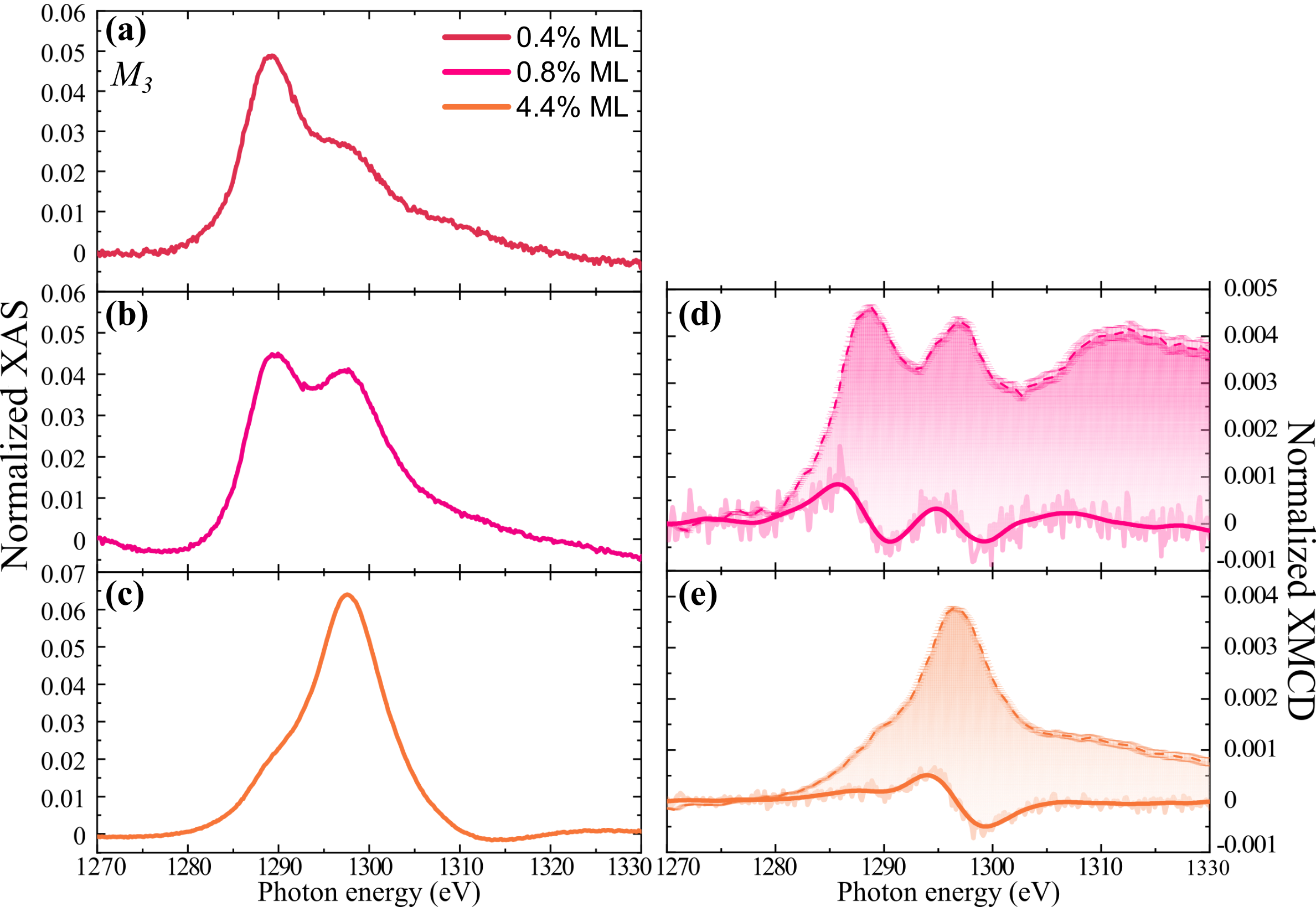}
        \caption{Normalized XAS at the $M_{3}$ edges for Nd on HOPG are presented for all prepared coverages (a, b, c). Normalized and integrated XMCD spectra at the $M_{3}$ edge ($\cfrac{p_{M_{3}}}{t_{M_{2,3}}}$ integral in Tab.~\ref{tab_pqt}) for a mixture of individual atoms and clusters (d) and ensemble of clusters (in majority) (e) are depicted. All spectra are normalized to the integral of the $M_{2,3}$ edge ($t_{M_{2,3}}$-integral in Tab.~\ref{tab_pqt}). Raw XMCD signals, displayed in transparent lines, are smoothed using a Sawatzky-Golay method (full lines) for enhanced clarity. The error bar is represented by the deviation around the central line of the XMCD integral (dashed lines).}
        \label{fig:SI3}
\end{figure}

\begin{figure}[h!]
        \centering \includegraphics[width=\textwidth]{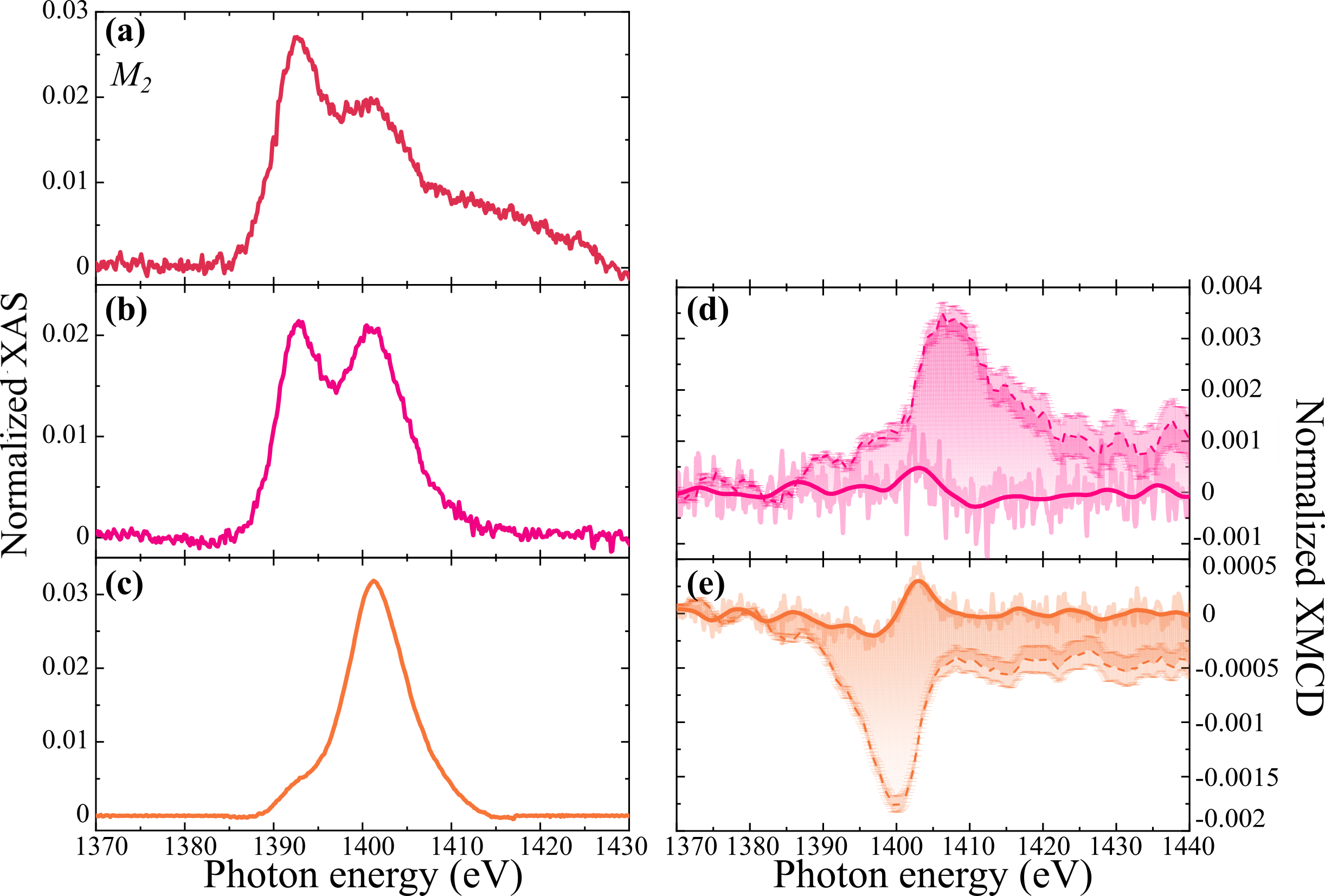}
        \caption{Normalized XAS at the $M_{2}$ edges for Nd on HOPG are presented for all prepared coverages (a, b, c). Normalized and integrated XMCD spectra at the $M_{2}$ edge ($\cfrac{q_{M_{2,3}}-p_{M_{3}}}{t_{M_{2,3}}}$ integral in Tab.~\ref{tab_pqt}) for a mixture of individual atoms and clusters (d) and ensemble of clusters (in majority) (e) are depicted. All spectra are normalized to the integral of the $M_{2,3}$ edge ($t_{M_{2,3}}$-integral in Tab.~\ref{tab_pqt}). Raw XMCD signals, displayed in transparent lines, are smoothed using a Sawatzky-Golay method (full lines) for enhanced clarity. The error bar is represented by the deviation around the central line of the XMCD integral (dashed lines).}
        \label{fig:SI4}
\end{figure}

\begin{figure}[h!]
        \centering \includegraphics[width=\textwidth]{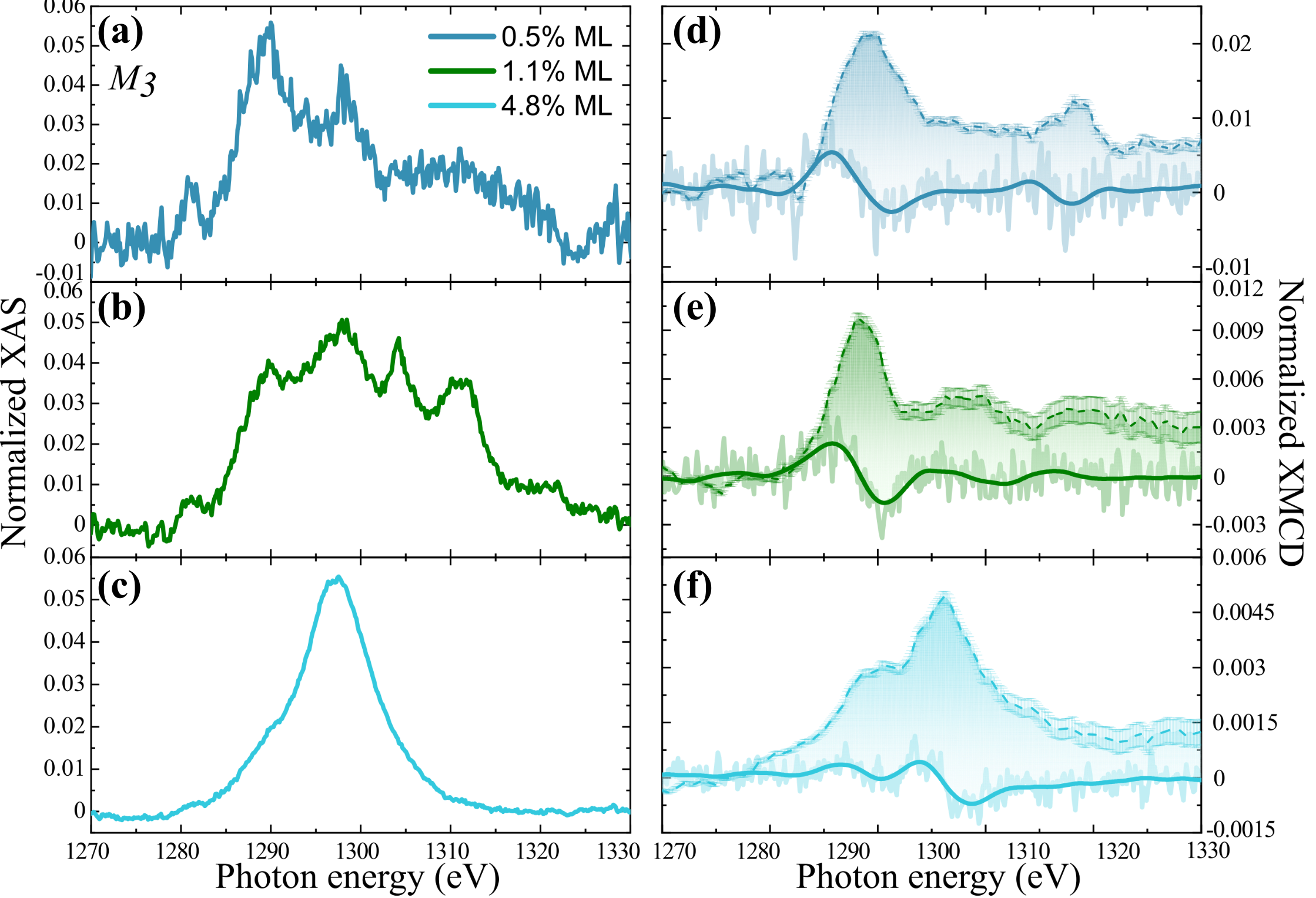}
        \caption{Normalized XAS at the $M_{3}$ edges for Nd on Ag(100) are presented for all prepared coverages (a, b, c). Normalized and integrated XMCD spectra at the $M_{3}$ edge ($\cfrac{p_{M_{3}}}{t_{M_{2,3}}}$ integral in Tab.~\ref{tab_pqt}) for an ensemble of individual atoms (d), a mixture of individual atoms and clusters (e) and ensemble of clusters (e) are depicted. All spectra are normalized to the integral of the $M_{2,3}$ edge ($t_{M_{2,3}}$-integral in Tab.~\ref{tab_pqt}). Raw XMCD signals, displayed in transparent lines, are smoothed using a Sawatzky-Golay method (full lines) for enhanced clarity. The error bar is represented by the deviation around the central line of the XMCD integral (dashed lines).}
        \label{fig:SI5}
\end{figure}

\begin{figure}[h!]
        \centering \includegraphics[width=\textwidth]{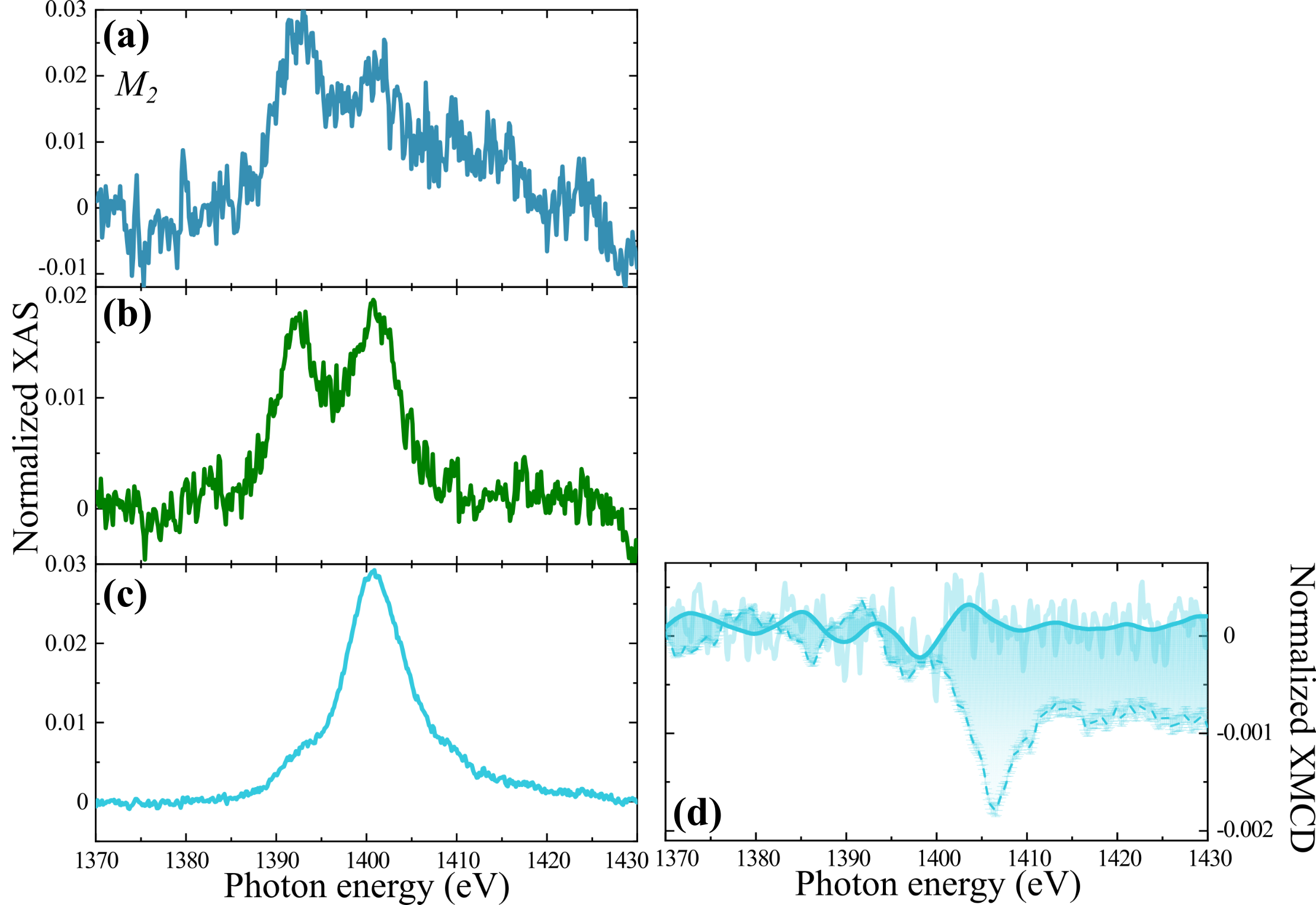}
        \caption{Normalized XAS at the $M_{2}$ edges for Nd on Ag(100) are presented for all prepared coverages (a, b, c). Normalized and integrated XMCD spectrum at the $M_{2}$ edge ($\cfrac{q_{M_{2,3}}-p_{M_{3}}}{t_{M_{2,3}}}$ integral in Tab.~\ref{tab_pqt}) for ensemble of clusters (d) is depicted. All spectra are normalized to the integral of the $M_{2,3}$ edge ($t_{M_{2,3}}$-integral in Tab.~\ref{tab_pqt}). Raw XMCD signal, displayed in transparent lines, is smoothed using a Sawatzky-Golay method (full lines) for enhanced clarity. The error bar is represented by the deviation around the central line of the XMCD integral (dashed lines).}
        \label{fig:SI6}
\end{figure}

\begin{figure}[h!]
        \centering \includegraphics[width=\textwidth]{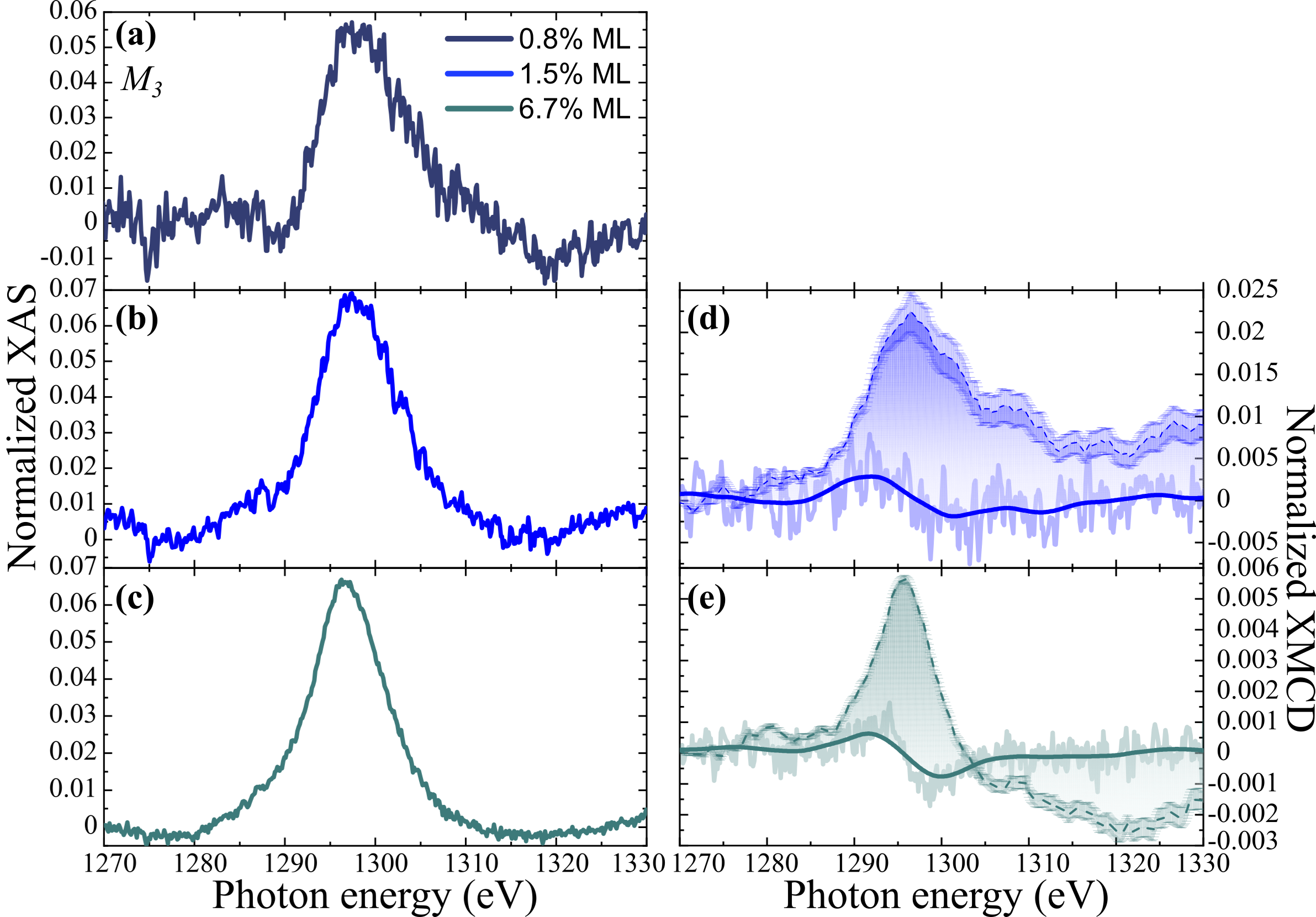}
        \caption{Normalized XAS at the $M_{3}$ edges for Nd on Pb(111) are presented for all prepared coverages (a, b, c). Normalized and integrated XMCD spectra at the $M_{3}$ edge ($\cfrac{p_{M_{3}}}{t_{M_{2,3}}}$ integral in Tab.~\ref{tab_pqt}) for a mixture of individual atoms and clusters (d) and ensemble of clusters (in majority) (e) are depicted. All spectra are normalized to the integral of the $M_{2,3}$ edge ($t_{M_{2,3}}$-integral in Tab.~\ref{tab_pqt}). Raw XMCD signals, displayed in transparent lines, are smoothed using a Sawatzky-Golay method (full lines) for enhanced clarity. The error bar is represented by the deviation around the central line of the XMCD integral (dashed lines).}
        \label{fig:SI7}
\end{figure}

\begin{figure}[h!]
        \centering \includegraphics[width=\textwidth]{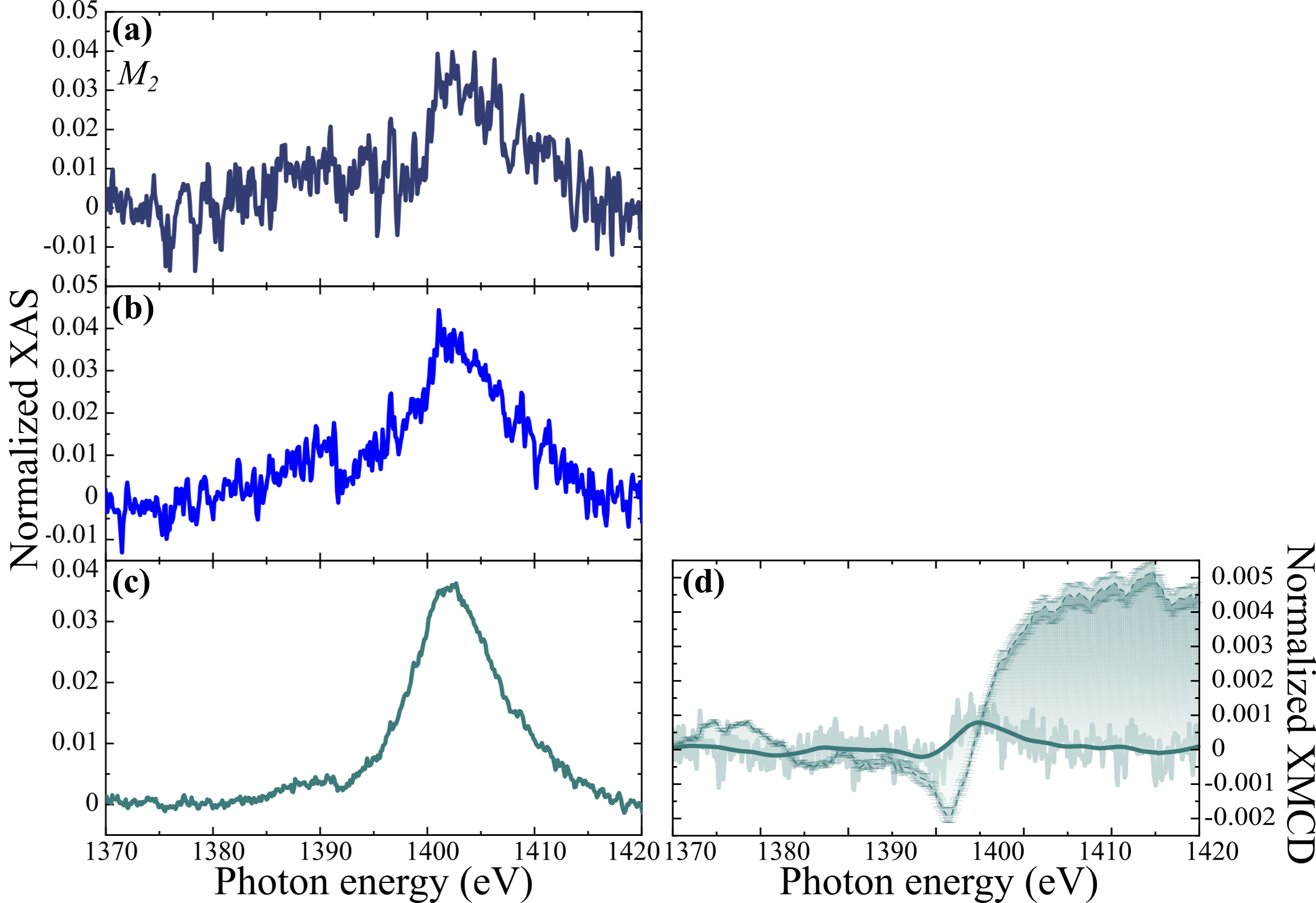}
        \caption{Normalized XAS at the $M_{2}$ edges for Nd on Pb(111) are presented for all prepared coverages (a, b, c). Normalized and integrated XMCD spectrum at the $M_{2}$ edge ($\cfrac{q_{M_{2,3}}-p_{M_{3}}}{t_{M_{2,3}}}$ integral in Tab.~\ref{tab_pqt}) for ensemble of clusters (d) is depicted. All spectra are normalized to the integral of the $M_{2,3}$ edge ($t_{M_{2,3}}$-integral in Tab.~\ref{tab_pqt}). Raw XMCD signal, displayed in transparent lines, is smoothed using a Sawatzky-Golay method (full lines) for enhanced clarity. The error bar is represented by the deviation around the central line of the XMCD integral (dashed lines).}
        \label{fig:SI8}
\end{figure}

\begin{figure}[h!]
        \centering \includegraphics[width=\textwidth]{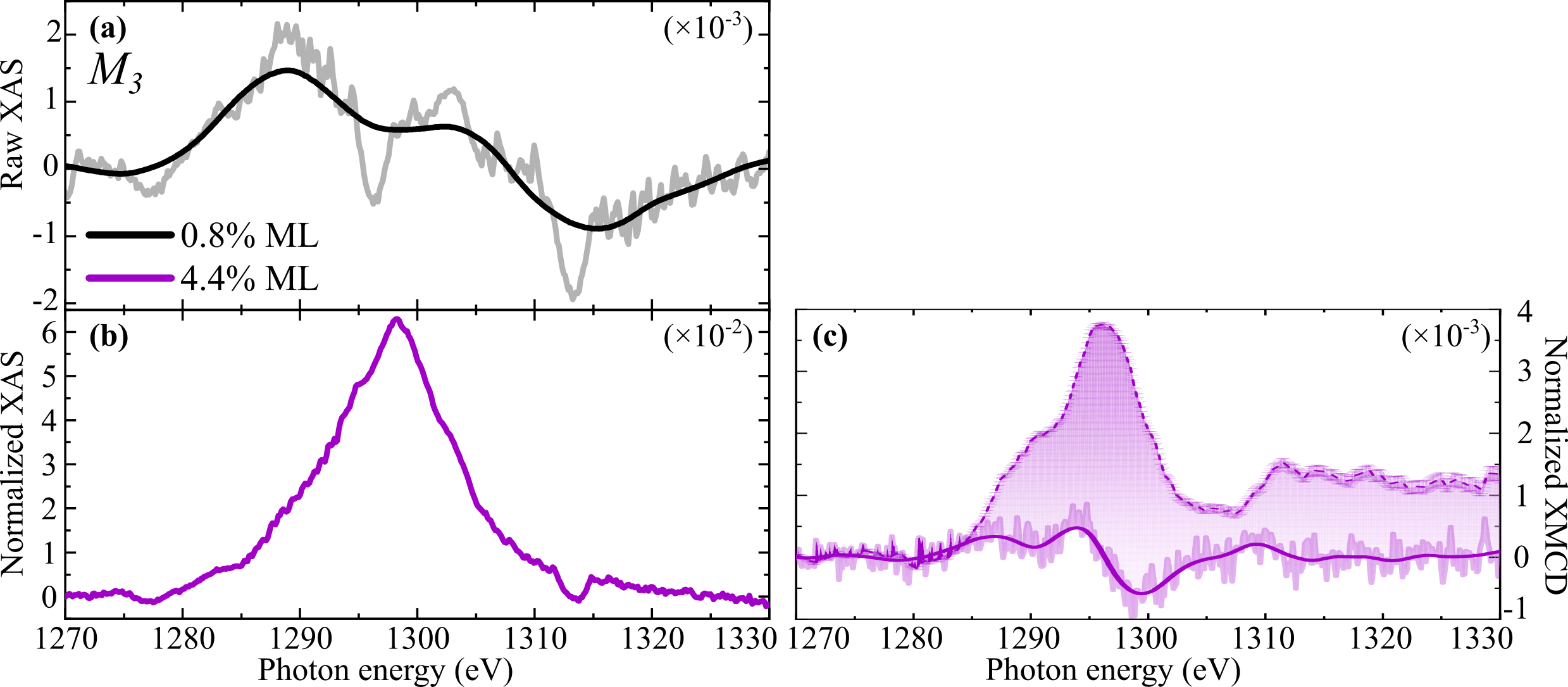}
        \caption{XAS at the $M_{3}$ edges for Nd on Gr/Cu(111) are presented for all prepared coverages. (a) Raw XAS is presented due to the strong Gr/Cu(111) background at this edge. (b) Normalized XMCD at the $t_{M_{2,3}}$ integral. Normalized and integrated XMCD spectrum at the $M_{3}$ edge ($\cfrac{p_{M_{3}}}{t_{M_{2,3}}}$ integral in Tab.~\ref{tab_pqt}) for ensemble of clusters (d) is depicted. Raw XAS and XMCD signals, displayed in transparent lines, are smoothed using a Sawatzky-Golay method (full lines) for enhanced clarity. The error bar is represented by the deviation around the central line of the XMCD integral (dashed lines).}
        \label{fig:SIGrCuM3}
\end{figure}

\begin{figure}[h!]
        \centering \includegraphics[width=\textwidth]{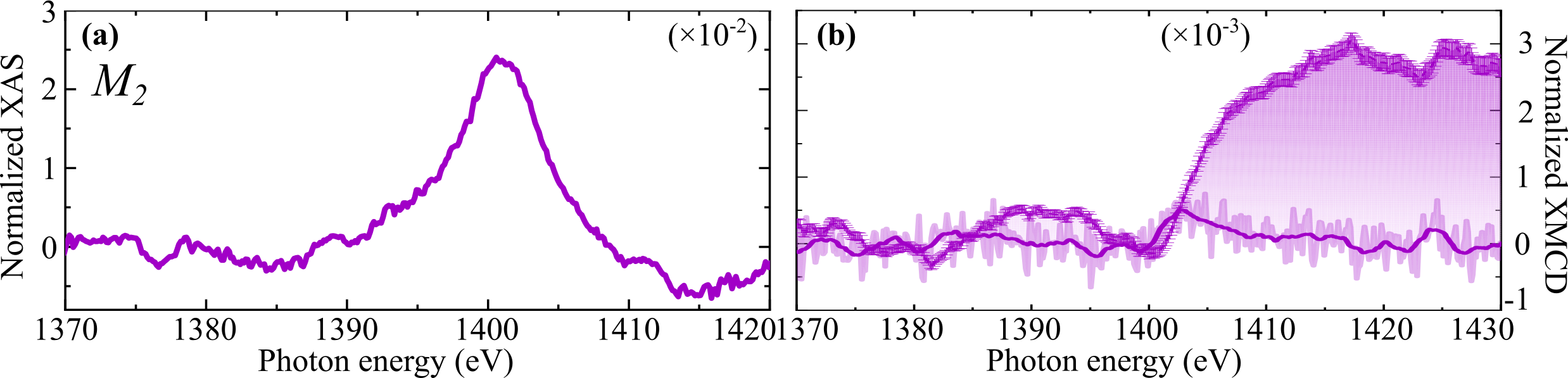}
        \caption{Normalized XAS at the $M_{2}$ edge for Nd on Gr/Cu(111) is presented for 4.4\%~ML (a). Normalized and integrated XMCD spectrum at the $M_{2}$ edge ($\cfrac{q_{M_{2,3}}-p_{M_{3}}}{t_{M_{2,3}}}$ integral in Tab.~\ref{tab_pqt}) for ensemble of clusters (d) is depicted. All spectra are normalized to the integral of the $M_{2,3}$ edge ($t_{M_{2,3}}$-integral in Tab.~\ref{tab_pqt}). Raw XMCD signal, displayed in transparent lines, is smoothed using a Sawatzky-Golay method (full lines) for enhanced clarity. The error bar is represented by the deviation around the central line of the XMCD integral (dashed lines).}
        \label{fig:SIGrCuM2}
\end{figure}

To compare the acquired XAS and XMCD spectra for different coverages on different substrates (Figs.~\ref{fig:SI3} to \ref{fig:SIGrCuM2}), the following analysis was carried out. First, the XAS($\mu^{+}$) and XAS($\mu^{-}$) spectra (Fig.~\ref{fig:SIDT}a, solid lines) were adjusted by subtracting an averaged background profile (Fig.~\ref{fig:SIDT}a, dotted lines). These polarization-dependent background profiles were obtained before Nd deposition on the pristine samples, ensuring no other experimental parameters were altered. Second, we summed the subtracted XAS($\mu^{+}$) and XAS($\mu^{-}$) to obtain the XAS, as shown in Fig.~\ref{fig:SIDT}b (cyan). This resulting spectrum exhibits a step-like feature (Fig.~\ref{fig:SIDT}b, blue), which originates from non-resonant excitations, i.e., transitions from the core state to the continuum. The step profile is described by the following function:
\[
f(x) = a + bx + cx^{2} + d \left( 1 - \frac{1}{1 + \exp{\left( \frac{x - e}{w} \right)}} \right)
\]
where \( x \) is the photon energy range, \( a \), \( b \), \( c \), \( c \) and \( d \) are fitting parameters, and \( w \) is the step width, set to 4~eV for the \( M_{2,3} \) edges. This process resulted in an almost featureless background across all photon energies, except at the position of the respective absorption peak. The remaining oscillatory profile observed after the peak is attributed to extended X-ray absorption fine structure. By repeating this procedure for the \( M_{2} \) edge, the integral of the XAS whiteline, referred to as \( t_{M_{2,3}} \), can be calculated. This allows for the normalization of the acquired spectra, as demonstrated throughout this manuscript. From the $M_{2,3}$ XAS measurements discussed in the main text (Fig.~2) and illustrated in Figs.~\ref{fig:SI3} to \ref{fig:SIGrCuM2}, we observe a substrate-dependent energy shift of the whole spectrum, attributed to changes in the occupation of the $4f$ (only on HOPG, Ag(100) and Gr/Cu(111)) and $5d$ shells.

From the XMCD measurements acquired in the direction of the easy-magnetization axis for the lowest coverage, we observe that the $M_{3}$ edge exhibits a characteristic peak-derivative shape for Nd atoms, which reverses its sign at the $M_{2}$ edge, as discussed in the main text. Unlike the $L_{2,3}$ spectra~\cite{van_veenendaal_branching_1997, parlebas_x-ray_2006}, the $M_{2,3}$ XMCD is not dominated by the so-called \textit{breathing effect}~\cite{van_veenendaal_branching_1997, parlebas_x-ray_2006} induced by the $4f$-$5d$ exchange interaction~\cite{harmon_spin-polarized_1974}. This is particularly true for Nd compared to the situation observed in Gd~\cite{singha_mapping_2021}, due to the reduced $4f$-$5d$ exchange interaction~\cite{fukui_x-ray_2001}. Furthermore, the characteristic peak-derivative of the $M_{2,3}$ XMCD, and the reversal in sign between transitions from the $3p_{3/2}$ and $3p_{1/2}$ core states, indicate a very weak $5d$ spin-dependent orbital \textit{breathing} effect. This confirms that this effect does not contribute significantly to the XMCD signal~\cite{van_veenendaal_branching_1997, goedkoop_strong_1997}. Therefore, it is possible to apply spin and orbital sum rules~\cite{thole_x-ray_1992, carra_x-ray_1993} neglecting this spurious effect. The orbital and spin moment for a $5d$ final state can be calculated using the following sum rules~\cite{thole_x-ray_1992, carra_x-ray_1993}:
\begin{subequations}
\begin{equation}\label{eq:5dorb}
\mu_{L}^{5d} = -n_{h}^{5d} \cfrac{4q_{M_{2,3}}}{3t_{M_{2,3}}}
\end{equation}
\begin{equation}\label{eq:5dspin}
\mu_{S+D}^{5d} = -n_{h}^{5d} \cfrac{6p_{M_{3}}-4q_{M_{2,3}}}{t_{M_{2,3}}}
\end{equation}
\textrm{Consequently:}
\begin{equation}\label{eq:5dmag}
\mu_{5d} = -n_{h}^{5d} \cfrac{6p_{M_{3}}-\frac{8}{3}q_{M_{2,3}}}{t_{M_{2,3}}}
\end{equation}
\end{subequations}
With $p_{M_{3}}$ the integral of the XMCD signal over $M_{3}$ edge, $q_{M_{2,3}}$ the integral of the XMCD signal over $M_{3}$ + $M_{2}$ edges, $n_{h}^{5d}$ the number of the $5d$ holes in the ground state and $\mu_{5d}$ the magnetic moment on the $5d$ orbital. The choice of using $\mu_{5d}$ is motivated by the fact that the contribution of the $q_{M_{2,3}}$, affected by the low-signal to noise at the $M_{2}$, is less relevant than for $\mu_{L}^{5d}$ and $\mu_{S+6D}^{5d}$. All these values are summarized in Tab.~\ref{tab_pqt}. 

\begin{table*}[htbp]
\hspace{-3cm}
\caption{Values of the integrals at the $M_{3}$ ($p_{M_{3}}$), $M_{2}$ ($q_{M_{2,3}}-p_{M_{3}}$), and the calculated magnetic moment for the $5d$ orbitals ($\mu_{5d}$ in $\mu_{B}$) on HOPG (Figs.~\ref{fig:SI3}-\ref{fig:SI4}), Ag(100) (Figs.~\ref{fig:SI5}-\ref{fig:SI6}), Pb(111) (Figs.~\ref{fig:SI7}-\ref{fig:SI8}), and Gr/Cu(111) (Figs.~\ref{fig:SIGrCuM3}-\ref{fig:SIGrCuM2}). The values of $p_{M_{3}}$ and $q_{M_{2,3}}-p_{M_{3}}$ are normalized by the value of the XAS integral at the $M_{2,3}$ edges ($t_{M_{2,3}}$) and are shown in $\times 10^{-3}$ unit. Missing values are replaced with zero in our calculations for the sum rules.}
\label{tab_pqt}
\hspace{-1.5cm}\begin{tabular*}{1.1\linewidth}{@{\extracolsep{\fill}} *{4}{S[table-format=1.4]} *{4}{S[table-format=1.4]} *{4}{S[table-format=1.4]} @{}}
\toprule
\toprule
\multicolumn{4}{c}{HOPG} & \multicolumn{4}{c}{Ag(100)} & \multicolumn{4}{c}{Pb(111)} \\
\cmidrule(lr){1-4} \cmidrule(lr){5-8} \cmidrule(l){9-12}
{$t_{M_{2,3}}$} & {$\cfrac{p_{M_{3}}}{t_{M_{2,3}}}$} & {$\cfrac{q_{M_{2,3}}-p_{M_{3}}}{t_{M_{2,3}}}$} & {$\mu_{5d}~(\mu_{B})$} & {$t_{M_{2,3}}$} & {$\cfrac{p_{M_{3}}}{t_{M_{2,3}}}$} & {$\cfrac{q_{M_{2,3}}-p_{M_{3}}}{t_{M_{2,3}}}$} & {$\mu_{5d}~(\mu_{B})$} & {$t_{M_{2,3}}$} & {$\cfrac{p_{M_{3}}}{t_{M_{2,3}}}$} & {$\cfrac{q_{M_{2,3}}-p_{M_{3}}}{t_{M_{2,3}}}$} & {$\mu_{5d}~(\mu_{B})$} \\
\midrule
0.234 & {---} & {---} & {---} & 0.017 & 8.6 & {---} & -0.27\pm0.09 & 0.017 & {---} & {---} & {---} \\
0.384 & 3.4 & 1.02 & -0.08\pm0.06 & 0.039 & 4.0 & {---} & -0.13\pm0.06 & 0.030 & 7.0 & {---} & -0.22\pm0.07 \\
1.460 & 0.7 & -0.46 & -0.03\pm 0.06 & 0.160 & 1.2 & -0.96 & -0.06\pm 0.07 & 0.129 & -1.8 & 4.54 & 0.17\pm0.07 \\
\bottomrule
\end{tabular*} 
\begin{tabular*}{0.33\linewidth}{@{\extracolsep{\fill}} *{4}{S[table-format=1.4]} @{}}
\multicolumn{4}{c}{Gr/Cu(111)} \\
\cmidrule(l){1-4}
{$t_{M_{2,3}}$} & {$\cfrac{p_{M_{3}}}{t_{M_{2,3}}}$} & {$\cfrac{q_{M_{2,3}}-p_{M_{3}}}{t_{M_{2,3}}}$} & {$\mu_{5d}~(\mu_{B})$} \\
\midrule
0.204 & 1.3 & 2.69 & -0.03\pm0.05 \\
\bottomrule
\bottomrule
\end{tabular*}
\end{table*}
A few remarks should be made about the values deduced from the sum rules:
\begin{itemize}
    \item Similar to the case of the sum rules on the $4f$ shells, and specifically for the intermediate and final coverage, here we are measuring the average signal, i.e. the average magnetic moment of all the species that coexist on the surface. 
    \item The number of holes is assumed to be 9.5, as an approximated intermediate value between the various configurations that could coexist and the ensemble, based on DFT results and estimation from the multi-orbital charge transfer model.
    \item The error bar on $\mu_{5d}$ reflects the square-root of the sum of the squares of the relative errors on the number of holes ($n_{h}^{5d} = 9.5 \pm 0.5$), on the integral $t_{M_{2,3}}$ and on the integrals of the $M_{3}$ and $M_{2}$ edges, represented by the deviation around the central line of the integral of the XMCD spectra.  

\end{itemize}
\newpage
\begin{center}
\textbf{\scalebox{1.5}{Nd/HOPG: $M_{3}$ edge}}
\end{center}

\newpage
\begin{center}
\textbf{\scalebox{1.5}{Nd/HOPG: $M_{2}$ edge}}
\end{center}

\newpage
\begin{center}
\textbf{\scalebox{1.5}{Nd/Ag(100): $M_{3}$ edge}}
\end{center}

\newpage
\begin{center}
\textbf{\scalebox{1.5}{Nd/Ag(100): $M_{2}$ edge}}
\end{center}

\newpage
\begin{center}
\textbf{\scalebox{1.5}{Nd/Pb(111): $M_{3}$ edge}}
\end{center}

\newpage
\begin{center}
\textbf{\scalebox{1.5}{Nd/Pb(111): $M_{2}$ edge}}
\end{center}

\newpage
\begin{center}
\textbf{\scalebox{1.5}{Nd/Gr/Cu(111): $M_{3}$ edge}}
\end{center}

\newpage
\begin{center}
\textbf{\scalebox{1.5}{Nd/Gr/Cu(111): $M_{2}$ edge}}
\end{center}

\clearpage
\subsection{2.3 Multiplet calculations at the $M_{2,3}$ edges}
To confirm that the magnetic moment measured by XMCD is indeed due to the $5d$ contribution, we performed multiplet calculations on the free Nd atom using QUANTY code~\cite{Haverkort_2012}. The choice to simulate the free atom was motivated by the need for reasonable computation time to gain a qualitative understanding of the XMCD measurements. Introducing the crystal field would require the identification of the related parameters which is beyond the scope of this work. Here, and differently from~\cite{singha_mapping_2021}, the use of multiplet calculation is not aimed at finding the electronic configuration of the $5d$ and the $6s$ due to the metallic character of these orbitals, upon Nd adsorption on the metal substrates, as supported by DFT calculations (see Fig.~\ref{fig:PDOS}).

\begin{figure}[h!]
        \centering \includegraphics[width=0.95\textwidth]{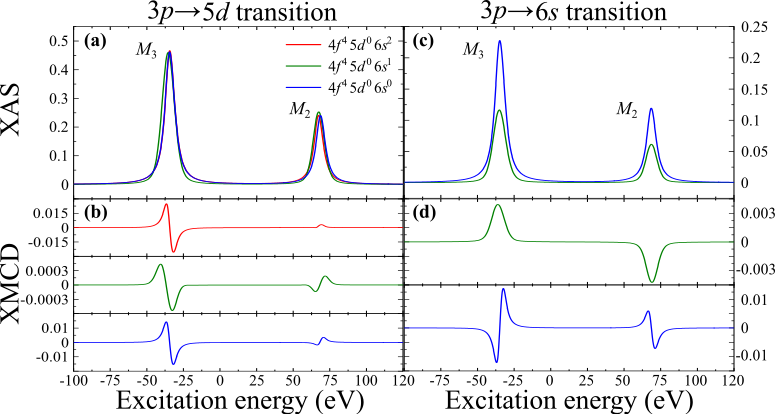}
        \caption{Multiplet calculations of the XAS (a, c) and XMCD (b, d) spectra at the $M_{2,3}$ edge of Nd free atom in the $4f^{4}5d^{0}$ configuration, with different occupations of the $6s^{k}$ orbitals ($k~=~0,~1~\textrm{and}~2$ in blue, green and red, respectively). The $3p \rightarrow 5d$ and $3p \rightarrow 6s$ transitions are shown in (a, b) and (c, d) respectively.}
        \label{fig:SI9}
\end{figure}

\begin{figure}[h!]
        \centering \includegraphics[width=0.95\textwidth]{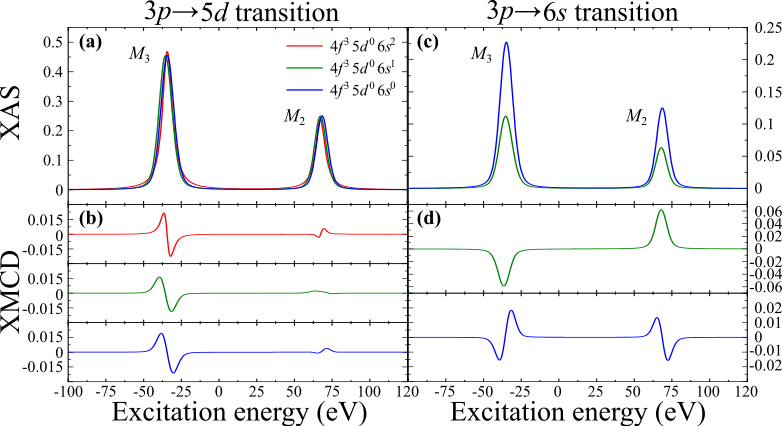}
        \caption{Multiplet calculations of the XAS (a, c) and XMCD (b, d) spectra at the $M_{2,3}$ edge of Nd free atom in the $4f^{3}5d^{0}$ configuration, with different occupations of the $6s^{k}$ orbitals ($k~=~0,~1~\textrm{and}~2$ in blue, green and red, respectively). The $3p \rightarrow 5d$ and $3p \rightarrow 6s$ transitions are shown in (a, b) and (c, d) respectively.}
        \label{fig:SI10}
\end{figure}

For the choice of the basis, we consider the associated Hilbert space defined by the electronic degeneracy of all the orbitals, i.e., $\binom{6}{n_{3p}} \times \binom{10} {n_{4f}}\times \binom{10}{n_{5d}} \times \binom{2}{n_{6s}}$, where $n_{3p} = 6$, $n_{4f} = 3~\textrm{or}~4$, $n_{5d} = 0$ and $n_{6s} = 0\textrm{,}~1~\textrm{or}~2$. Moreover, the calculation parameters are as follows: a) the Slater integrals $F^{0}$ produces a rigid shift of the spectrum, and do not play a role in the XAS shape, as the higher-order integrals $F^{k}$, and the $G^{k}$ terms (see Eq.~\eqref{U}) are taken from the Cowan code~\cite{cowan_theory_nodate}; b) a magnetic field of $B=6$~T, and a temperature of $T=2$~K, have been applied. Furthermore, since the $3p \rightarrow 5d$ and $3p \rightarrow 6s$ transitions are both dipole-allowed transitions, we separate the transition operator into two components, each reflecting a transition towards a different final state, to analyze their individual contributions to the XAS/XMCD spectrum. Figures~\ref{fig:SI9} and \ref{fig:SI10} show the transitions for different $4f$ occupancies, namely, $4f^{4}5d^{0}6s^{k}$ and $4f^{3}5d^{0}6s^{k}$, towards different final states, $3p \rightarrow 5d$ (a/b) and $3p \rightarrow 6s$ (b/c) on both figures. Two fairly simple and direct observations can be made from these results: firstly, the shape of the XMCD at the $M_{3}$ edge is reproduced only if we consider transitions to the $5d$ states; secondly, the shape of the XMCD spectrum of Nd on the $M_{3}$ edge is reproduced regardless of the $4f$ and $6s$ configuration considered. As conclusion, the $M_{2,3}$ XMCD indicates that the signal is dominated by $3p \rightarrow 5d$  transitions.

\clearpage
\section{3. Density functional theory calculations}
Density functional theory (DFT) calculations were done using the projector augmented-wave (PAW) method and planewave basis sets with the VASP code \cite{kresse_efficiency_1996,kresse_ultrasoft_1999}. As exchange and correlation functional we used the GGA approximation in its PBE version \cite{perdew_generalized_1996}. Missing van der Waals interactions in this functional were accounted for using Grimme's D3 method \cite{grimme_consistent_2010} with Becke-Johnson damping function \cite{grimme_effect_2011}. In order to set the size of the planewave basis set an energy cutoff of 500~eV was used for HOPG, while 400~eV was used for Ag and Pb. Surfaces were simulated using the slab method with 4 layers for HOPG and 5 layers for Ag(001) and Pb(111). The size of the periodic unit cell was $6\times6$ for HOPG and Ag(001), and $5\times3\sqrt{3}$ for Pb(111). A $(3\times3\times1)$ regular mesh was used for the sampling of the Brillouin zone. The position of all atoms except the two lower layers were relaxed until all forces were smaller than 0.01~eV/{\AA}.

For the treatment of Nd we have included the $4f$ electrons in the valence of the PAW potential. In order to improve the description of the correlations of these strongly localized electrons we have used the DFT+U method \cite{dudarev_electron-energy-loss_1998}. For the $U_{eff}=U-J$ parameter of Nd $4f$ we have used a value of 6.00~eV, following previous works \cite{kozub_electronic_2016,ma_neodymium_2022}. The PAW potential for Nd was generated in the $4f^{3}5d^{1}$ configuration, using 14 electrons as valence ($5s^{2}5p^{6}4f^{3}6s^{2}5d^{1}$). However, due to the good transferability of the potential we found solutions in both $4f^{3}$ and $4f^{4}$ configurations, depending on the environment of the Nd atom.

\begin{figure}[h!]
        \centering \includegraphics[width=\textwidth]{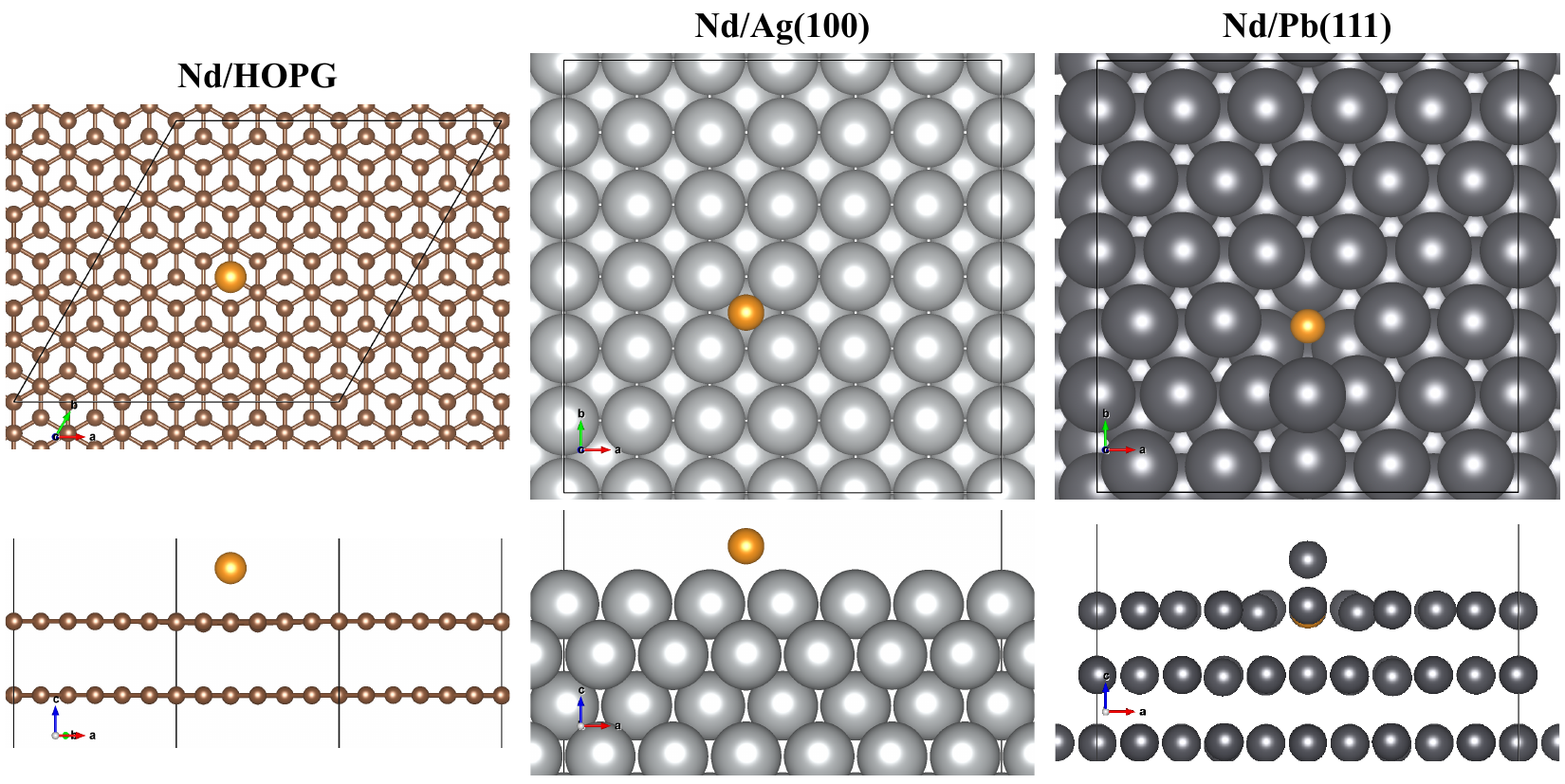}
        \caption{Relaxed geometries of the Nd atom on HOPG (left), Ag(100) (middle), and Pb(111) (right). Nd atom is shown in orange, while C, Ag and Pb atoms are shown in brown, light grey and dark grey, respectively. The unit cells are shown with black lines.}
        \label{fig:geom}
\end{figure}

We have considered different hollow, bridge and top adsorption sites of the Nd atom on all surfaces. In addition, in the case of Pb(111) we have also considered different subsurface configurations. For HOPG and Ag(100) the hollow adsorption site is energetically preferred (Fig.~\ref{fig:geom}). In the case of Pb(111), the Nd atom prefers to be embedded under the surface (Fig.~\ref{fig:geom}), in line with previous works on transition metal atoms \cite{choi_mapping_2017,mier_two_2024}. In Tab.~\ref{tab:SI_qm} we show the $4f$ and $5d$ occupations and magnetic moments of Nd on the different surfaces. 

\begin{table*}[htbp]
\caption{Occupations ($q$) and magnetic moments ($\mu$) of $4f$ and $5d$ orbitals of the Nd atom on the different surfaces. Magnetic moments are given in Bohr magnetons ($\mu_B$).}
\label{tab:SI_qm}
\begin{ruledtabular}
\begin{tabular}{ ccc  cc}
 & $\textrm{q}_{4f}$ & $\textrm{q}_{5d}$ & $\mu_{4f}$ & $\mu_{5d}$ \\
 \midrule
HOPG & 3.98 & 0.24 & 3.97 & 0.09 \\
Ag(100) & 3.99 & 0.27 & 3.97 & 0.06 \\
Pb(111) & 3.05 & 0.83 & 3.02 & 0.17 \\
\end{tabular}
\end{ruledtabular}
\end{table*}

On HOPG and Ag(100) Nd shows a $4f^{4}$ configuration. On Pb(111) the configuration depends on the environment: when Nd is on the surface, it also shows a $4f^{4}$ atomic-like configuration; however, when Nd is embedded under the surface, $4f^{3}$ bulk-like configurations can be found, like the one shown in Tab.~\ref{tab:SI_qm}.

\begin{figure}[h!]
        \centering \includegraphics[width=\textwidth]{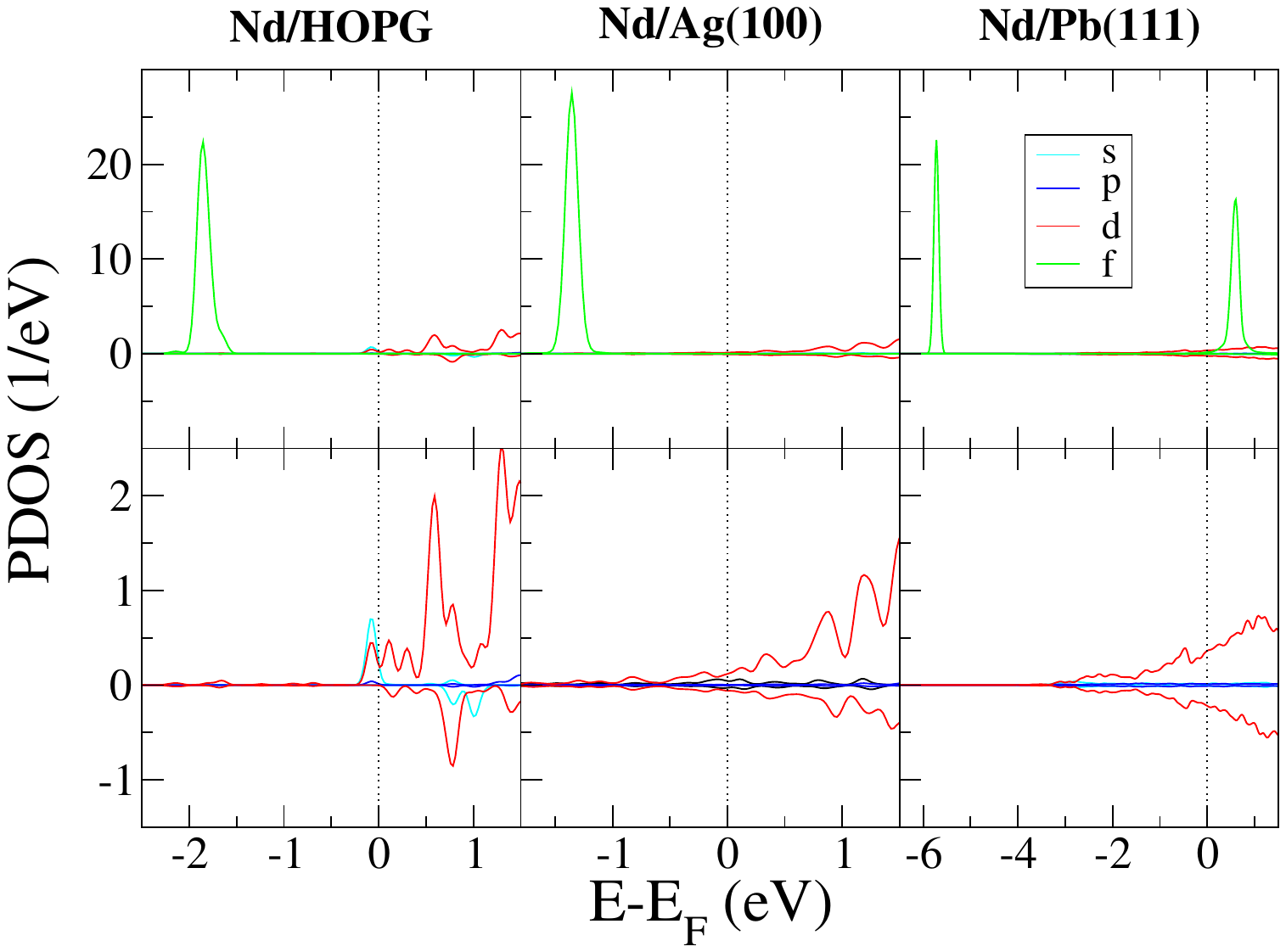}
        \caption{Projected density of states (PDOS) on the orbitals of the Nd atom on different surfaces. $s$, $p$, $d$, and $f$ orbitals are shown with cyan, blue, red and green lines, respectively. In the bottom row only $s$, $p$, and $d$ orbitals are shown. The dotted lines show the position of the Fermi energy.}
        \label{fig:PDOS}
\end{figure}

In Fig.~\ref{fig:PDOS} we show the projected density of states on the different surfaces. Here we can also see the different behavior on the Pb(111) surface compared to the other surfaces. On Pb(111) the Nd $4f$ electrons are much lower in energy, with some unoccupied $4f$ states close to the Fermi energy. The $5d$ states on that surface can be found down to -3~eV, showing the stronger hybridization with the Pb states.

\clearpage

%